\documentclass{aa} 

\usepackage{graphicx}
\usepackage{txfonts}
\usepackage[dvipsnames]{xcolor}
\usepackage{multirow}

\usepackage{subcaption}
\usepackage{caption}
\colorlet{lightmagenta}{magenta!60}

\usepackage{amsmath,amssymb}

\usepackage{ulem}

\defcitealias{BinneyTremaine2008}{BT08}
\def\BT08{\citetalias{BinneyTremaine2008}}

\usepackage{hyperref}
\hypersetup{
    colorlinks=true,
    linkcolor=black,
    citecolor=blue}
\newcommand{\orcid}[1]{\href{https://orcid.org/#1}{\includegraphics[height=10pt]{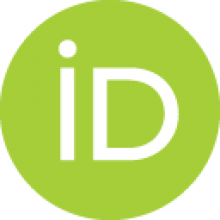}}}

\begin{document} 

   \title{Forming off-center massive black hole binaries in dwarf galaxies through Jacobi capture}

    \titlerunning{Forming off-center massive black hole binaries in dwarf galaxies through Jacobi capture}
    \authorrunning{T. L. François et al. }
    

   \author{Thibaut L. François
          \inst{1}\fnmsep\thanks{e-mail: thibaut.francois@proton.me}\orcid{0009-0001-0314-7038},
          Christian M. Boily
          \inst{1}\orcid{0000-0002-7274-6720},
          Jonathan Freundlich
          \inst{1}\orcid{0000-0002-5245-7796},
          Simon Rozier
          \inst{2}\orcid{0000-0001-6994-6708}
          \and
          Karina Voggel
          \inst{1}\orcid{0000-0001-6215-0950}
          }

   \institute{Université de Strasbourg, CNRS, Observatoire Astronomique de Strasbourg, UMR 7550, F-67000 Strasbourg, France
         \and
        Institute for Astronomy, University of Edinburgh, Royal Observatory, Blackford Hill, Edinburgh EH9 3HJ, UK
             }

   \date{Received 13 November 2023 / Accepted 26 April 2024}

 
  \abstract
  {It is well established that black holes reside in the central regions of virtually all types of known galaxies. Recent observational and numerical studies however challenge this picture, suggesting that intermediate-mass black holes in dwarf galaxies may be found on orbits far from the center. In particular, constant-density cores minimize orbital energy losses due to dynamical friction, and allow black holes to settle on stable off-center orbits. Using controlled simulations, we study the dynamics of off-center black holes in dwarf galaxies with such cores. We propose a new scenario to describe off-center mergers of massive black holes, starting with a Jacobi capture. We focus on initially circular and co-planar black hole orbits and explore a large parameter space of black hole masses and orbital parameters. We find that Jacobi captures are a complex and chaotic phenomenon that occurs in about $13\%$ of cases in this simplified setup, and we quantify how the likelihood of capture depends on the simulation parameters. We note that this percentage is likely an upper limit of the general case. Nevertheless, we show that Jacobi captures in cored dwarf galaxies can facilitate the formation of off-center black hole binaries, and that this process is sufficiently common to have a substantial effect on black hole growth. While our setup only allows for temporary captures, we expect dissipative forces from baryons and post-Newtonian corrections to maintain the captures over time and to lead to the formation of stable binary systems. This motivates future studies of the effectiveness of such dissipative forces, within stripped nuclei or globular clusters, in forming stable bound systems.}



   \maketitle
%

\section{Introduction}

Over the last decades, observations have uncovered massive black holes (BHs) at the center of massive galaxies \citep[see ][]{Kormendy95}. Their mass correlates with the properties of the host galaxy \citep[][]{Ferrarese2000,Gebhardt2000,vanDenBosch2016}, which suggests a common past history. Recent studies with different probes (e.g., optical, X-ray, Radio or infrared) have brought to light the presence of such BHs within dwarf galaxies \citep[see ][]{Reines2013, Pardo2016, Mezcua2018, Mezcua2019, Zaw2020, Cann2020}. And although a clear consensus has yet to emerge due to limitations in observed kinematics and dynamical modeling, studies reveal indications of intermediate-mass black holes (IMBHs) at the centers of some globular clusters \citep[see ][]{Noyola2008, vanDerMarel2010, McNamara2012, Lutzgendorf2013a, Lutzgendorf2013b, Lanzoni2013}. Major galaxy mergers, accretion events and in-spiraling of globular clusters can lead to multiple of these BHs within the same galaxy \citep[see][]{Schweizer2018, Nguyen2019, Kollatschny2020, Shen2021, Stemo2021, Voggel2022, Pechetti2022}.\\

\begin{figure*}
    \centering
    \includegraphics[width=1.7\columnwidth]{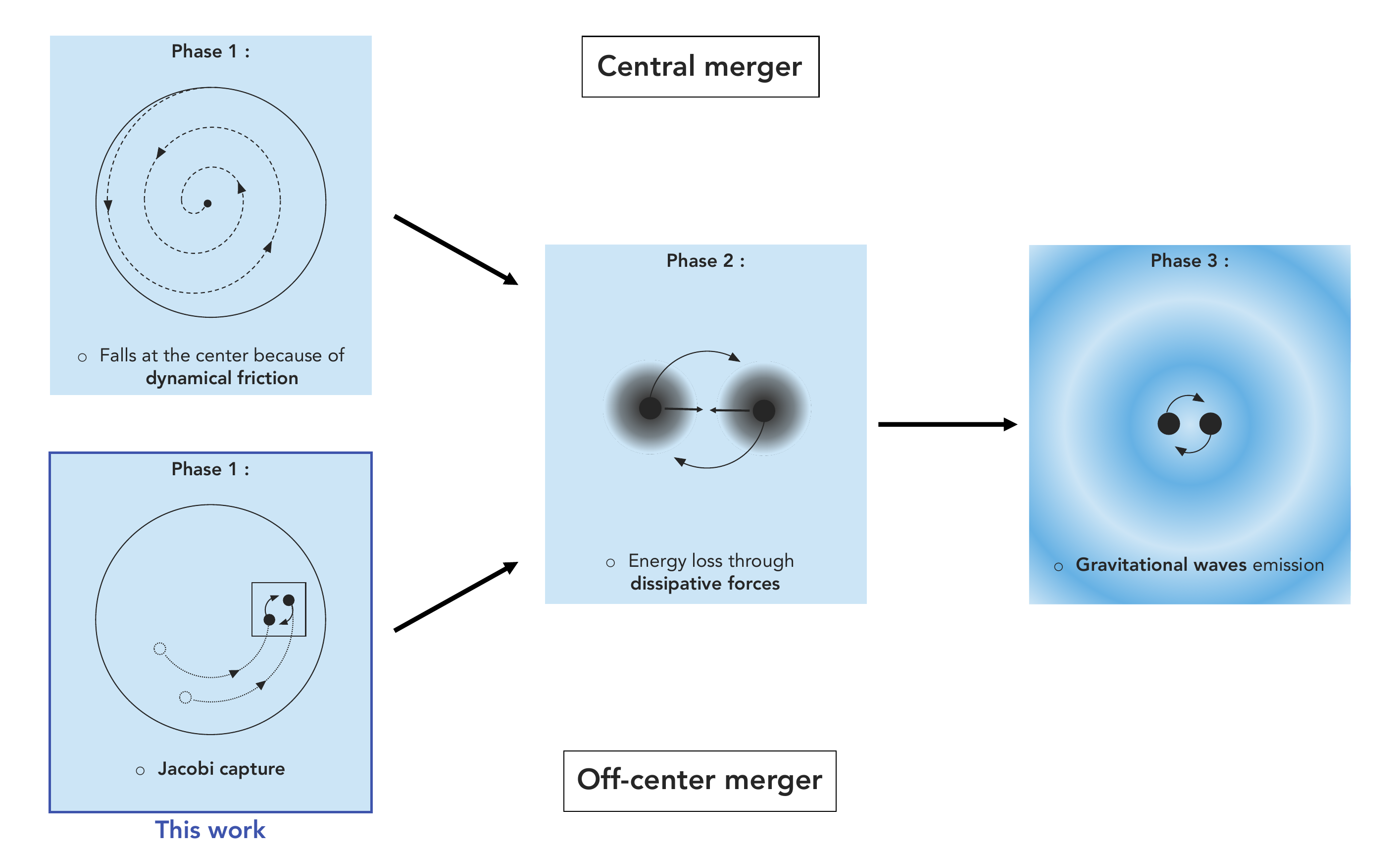}
    \caption{Central versus off-center BH merger scenarios. Top row: classical massive BH merger scenario in the galactic center. Phase 1: the massive BHs lose angular momentum through dynamical friction with the galaxy and halo, which causes them to sink toward the center. Phase 2: once both BHs have reached the center, they may form a bound pair and continue to lose energy via dissipative forces. Phase 3: the binary eventually enters the relativistic regime and loses energy through gravitational waves emission, allowing the two BHs to merge. Bottom row: proposed scenario for an off-center merger. Phase 1: the BHs are on off-center orbits e.g., as a result of long dynamical time for their radius to decrease through dynamical friction or core stalling acting against dynamical friction. If both BHs are sufficiently close energetically, they may form an off-center bound pair through Jacobi capture. This is the scenario explored in the present paper.}
    \label{fig-schemeMerger}
\end{figure*}

These BHs can then merge through a three-stage process \citep[][]{Begelman1980}: first, massive BHs are driven to the center of the galaxy, due to the loss of angular momentum from dynamical friction (\citealt[][]{Chandrasekhar43}; \citealt[][hereafter \citetalias{BinneyTremaine2008}]{BinneyTremaine2008}, Sect. 8.1, Eq. 8.1-4). When the separation between two BHs is comparable to the gravitational influence radius of the larger one, they form a bound pair and a second phase takes place. The binary gives away some of its angular momentum via three-body interactions to stars with lower angular momentum \cite[see][]{Heggie75}. In scenarios where a sufficient number of stars are available to facilitate angular momentum exchange, the binary system enter the relativistic regime, emitting gravitational waves before merging. These three distinct phases constitute the conventional framework for the merger of massive BHs through the assembly of galaxies (depicted in the top panel of Fig. \ref{fig-schemeMerger}).\\

The merging scenario relies mainly on dynamical friction to bring BHs together at the center. However, the time required to bring IMBH ($10^3 - 10^6 M_\odot$) to the center of a typical dwarf galaxy can be much longer than a Hubble time in the case of cored dark matter density profiles and/or low stellar densities\footnote{With Eq. 8.1 of \BT08, we compute a time $\sim 22\ \rm Gyr$ for a $10^6 M_\odot$ BH to fall from 5 kpc to the center of a cored dwarf spheroidal galaxy with a virial mass of $10^{11} M_\odot$. For a $10^3 M_\odot$ BH falling from 10 kpc, it takes $\sim 85\ 000\ \rm Gyr$.} \citep[see][]{Bellovary2019, Pfister2019, Bellovary2021}. As a consequence, BHs are left on off-center orbits, with their guiding radius reducing only very slightly over 14 Gyr. According to simulations and observations, it is estimated that half of the IMBHs are off-center in dwarf galaxies \citep[see][]{Bellovary2019, Reines2020, Bellovary2021}. Furthermore, their electromagnetic detection is highly challenging due to their low accretion luminosities, and their dynamical detection is equally problematic as they have a very small impact on the dynamics of the stars within their sphere of influence.\\

In certain instances, however, the long infall time argument alone does not provide a satisfactory explanation for the existence of off-center objects, such as in the case of the Fornax dwarf galaxy. This galaxy harbors five globular clusters orbiting at approximately 1 kpc from its center \citep[]{Mackey2003}, and based on conventional assumptions, these clusters should have migrated from their current locations within a few billion years \citep[see ][though this picture has been challenged, given the modeling freedom of the initial GC formation radius, and uncertainties arising from projection effects, see e.g., \citealt{Borukhovetskaya2022}]{Goerdt2006}. This is, in part, due to dynamical processes which cannot be modeled using Chandrasekhar dynamical friction alone: numerical simulations have revealed instances of the perturber's fall coming to a halt, known as "core stalling" \citep[see][]{Read2006, Inoue2011, Petts2015}, as well as the occurrence of an outward push, or "dynamical buoyancy" \citep[see][]{Cole2012}. These phenomena are observed in constant-density cores \citep[as those observed at the center of dark matter halos in low-mass galaxies, see ][]{BinneyEvans2001, Borriello2001, deBlok2001}, and are not accounted for in Chandrasekhar’s description. Recent analytical models provide an explanation in terms of resonances \citep[see ][]{Boily2008, Kaur2018, Banik2021}: when a perturber enters the core region, significant resonances responsible for dynamical friction are suppressed, leading to a cessation in the fall. Within a core, these resonances can even give energy to the perturber, causing it to migrate outward.\\

In light of these cases where IMBHs are found off-center, either because their infall time greatly exceeds the dynamical time of the system or because it is halted due to the presence of a constant density core, the question arises whether a IMBH merger is possible outside the galactic center. Indeed, the aforementioned three-phase scenario must be revised. Firstly, there is no guarantee that the BHs may get sufficiently close in phase space to form a binary, as opposed to the classical case where the galactic center acts as an attractor, allowing an agglomeration at a single point. In this context, specific phase-space conditions must be met for them to form bound configurations away from the center. 

Our proposed scenario of off-center Jacobi capture may be modeled, as a first approximation, as a circular-restricted three-body problem. This is a variation of the three-body problem in which two very light objects orbit around a much more massive body. This problem, known as Hill's problem, has been extensively studied to describe the movements of satellites around a planet \citep[][]{PetitHenon86, HenonPetit86b, Petit90} as well as gravitational wave capture of stellar-mass BHs around a supermassive BH \citep[][]{Boekholt2023, Li2022}. The principal distinctions in our case lies in the fact that a galaxy is an extended massive body rather than point-like, so the mass of the central body depends on the orbital radius of the BHs and the shape of the potential is different, featuring a density core at the center. For a merger to occur, it is first necessary for the two light bodies to meet a suitable configuration where their gravitational attraction becomes comparable to the force generated by the massive body. 

In this case, they bond for a certain period of time, known as Jacobi capture, and produce several close encounters. This is the first assumed step in an off-center merger scenario. Then, it has been observed that dissipative forces play a crucial role in the hardening of the binary \citep[e.g.,][]{Fabian75, Goldreich2002}. In \citet{Li2022, Boekholt2023} this energy dissipation is facilitated by the emission of gravitational waves, while in \citet{Tagawa2018, Tagawa2020, Li2022, Rowan2023, DeLaurentiis2023}, it is dynamical friction with the surrounding gas that tightens the binary. We anticipate similar behaviors for the case of two IMBHs: the bound stellar populations that typically surround them can generate dynamical friction on the other BH, eventually leading to the last step of the classical scenario (see bottom panel on Fig. \ref{fig-schemeMerger}).\\

In this paper, we focus on the first step of this process: the Jacobi capture. We aim to analyze which initial configurations, in terms of BH masses and kinematics, can allow Jacobi capture to occur within a Hubble time. In order to isolate the capture step, we perform idealized simulations of two point-like BHs in a smooth background gravitational potential, varying the initial BH masses and kinematics. This work allows to pinpoint regions in phase space where Jacobi captures are most likely to occur, paving the way for a comprehensive investigation of the subsequent merger stages through $N$-body simulations.\\

This paper is organized as follows. In Section \ref{method}, we outline our methodology, including the gravitational potential, parameter selection, core stalling radius and influence radius computation, capture criterion and an examination of the chaotic behavior of Jacobi captures. In Section \ref{results}, we present our findings. Initially, we examine the general case involving five parameters, wherein no assumptions are made regarding the cause of the offset of BHs. Subsequently, we explore the scenario where BHs are positioned at their core stalling radius. Lastly, we determine the probability of capture by employing a Monte Carlo generation of merger trees to select BH masses. Our outcomes are summarized in Section \ref{discussion} and we discuss our results.

\section{Method}
\label{method}

\subsection{Numerical setup}

In this study, we aim to explore the influence of different initial properties of BHs on the occurrence of a Jacobi capture. We also aim to provide physical insights on the capture process itself, occurring off-center in an extended galactic potential. As a consequence, we are reduced to a number of simplifying assumptions. Because we want to sample a large parameter space of initial BH masses and kinematics, we cannot afford to simulate the host galaxy as a live $N$-body component. Indeed, the problem we are considering requires high computational accuracy along two lines. First, we need to resolve the dynamics of the BH binary, implying a high time resolution of the integration. Second, an extreme number of galactic particles is required to resolve processes such as dynamical friction\footnote{We simulate IMBHs of $10^{3}-10^{6} M_\odot$ in a DM halo of $10^{11} M_\odot$. In order to resolve dynamical friction, a mass gap of $\sim 1/1000$ is required between the galactic particles and the BHs. Consequently, the mass of the galactic particles must not exceed $1 M_\odot$. We therefore need to simulate $10^{11}$ particles which is well outside the technical capabilities at our disposal in terms of memory and computing time.}. As a workaround, we consider the galaxy as a background static potential, in a simplified setup that enables us to execute thousands of simulations within a reasonable timeframe.\\

\begin{figure}
    \centering
    \includegraphics[width=1.\columnwidth]{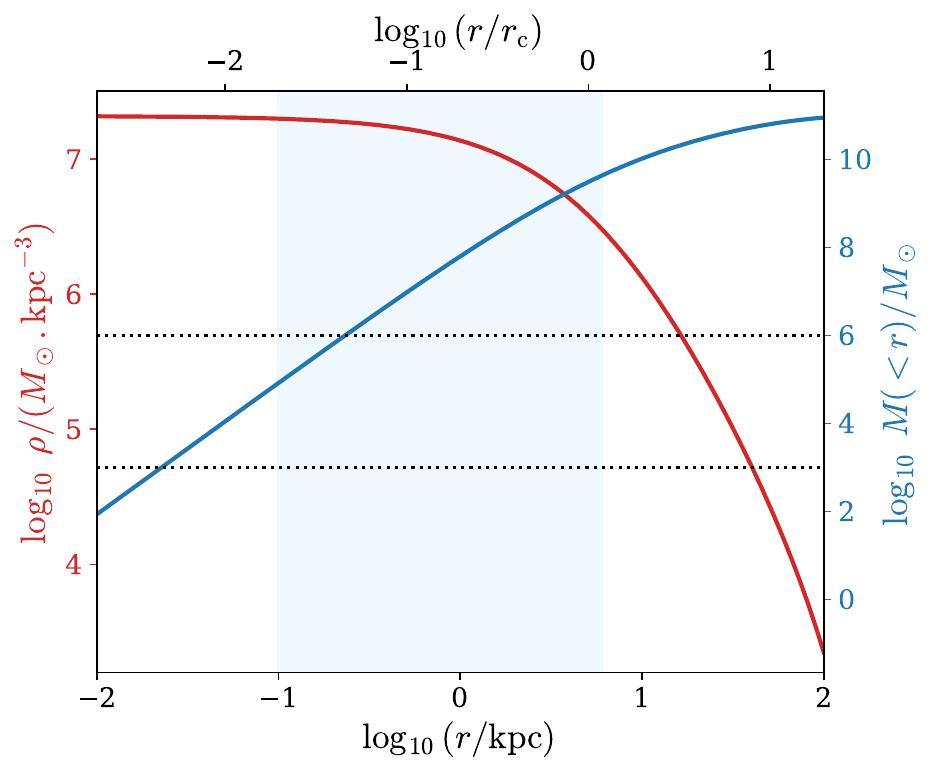}
    \caption{Density profile and integrated mass profile of the dark matter halo (background potential). The two horizontal lines represent the limits of the mass range we consider for the BHs. The shaded area denotes the scope of this study, and the $x$-axis at the top of the figure is expressed in units of the core radius of the potential $r_{\mathrm{c}}$, defined by its inflection point.}
    \label{fig-denProf}
\end{figure}

We integrate the orbits of two BHs in an external gravitational potential representing the dark matter halo of a dwarf galaxy. Simulations in this paper made use of the REBOUND N-body code \citep{rebound}. The simulations were integrated using IAS15, a 15th order Gauss-Radau integrator \citep{reboundias15}. This integrator incorporates a step-size control mechanism to automatically select an optimal timestep and is adept at handling close encounters and orbits with high eccentricity. With systematic errors maintained well below machine precision, IAS15 demonstrates adherence to Brouwer's law, where energy errors exhibit random walk behavior.\\
We choose a dwarf galaxy halo as various studies indicate that this type of galaxy hosts a substantial population of off-center BHs \citep[][]{Bellovary2019, Bellovary2021}. Furthermore, in the standard $\Lambda$CDM paradigm dwarf galaxies are dominated by dark matter, hence we can initially approximate them by considering only their halo. We consider a galaxy with a $10^{11} M_\odot$ cored dark matter halo with a virial radius $r_{200}=100\ \rm kpc$ (this corresponds to a stellar mass of $\sim 2\times 10^{8} M_\odot$). This halo mass range is expected to develop cored profiles more efficiently \citep[see ][]{Tollet2016, freundlich2020}. 
The dark matter component of our dwarf galaxy model is described by a cored ($\alpha, \beta, \gamma$) density profile \citep{Hernquist90, Zhao96} 
\begin{equation}
    \rho = \rho_0 \left( \frac{r}{a} \right)^{-\gamma} \left [ 1+\left( \frac{r}{a} \right)^{\alpha} \right ]^{\frac{\gamma - \beta}{\alpha}},
    \label{eq-densityProfile}
\end{equation}
with $\alpha = 1,\ \beta = 3$ and $\gamma = 0$ respectively the outer, transitional and inner slope of the density profile. The scale radius is $a = 6.7$ kpc. The density profile as well as the integrated mass profile are depicted in Fig. \ref{fig-denProf}.

\subsection{Toy model}

We adopted a simplified model where the two BHs are on circular orbits in the same plane. This configuration closely resembles Hill's problem \citep[][]{PetitHenon86, HenonPetit86b, Petit90, Boekholt2023}. However, our massive body is extended rather than point-like, so its mass depends on the orbital radius of the BHs, and the potential shape differs, featuring a density core at the center. If our BHs are close to the center (which is the case for core stalling), the mass of the central body is comparable to the mass of the BHs. When the proximity of the two BHs leads to gravitational forces comparable to those exerted by the galaxy, temporary captures occur. These are known as Jacobi captures and are marked by multiple close encounters between BHs.

\begin{figure}
    \centering
    \includegraphics[width=0.8\columnwidth]{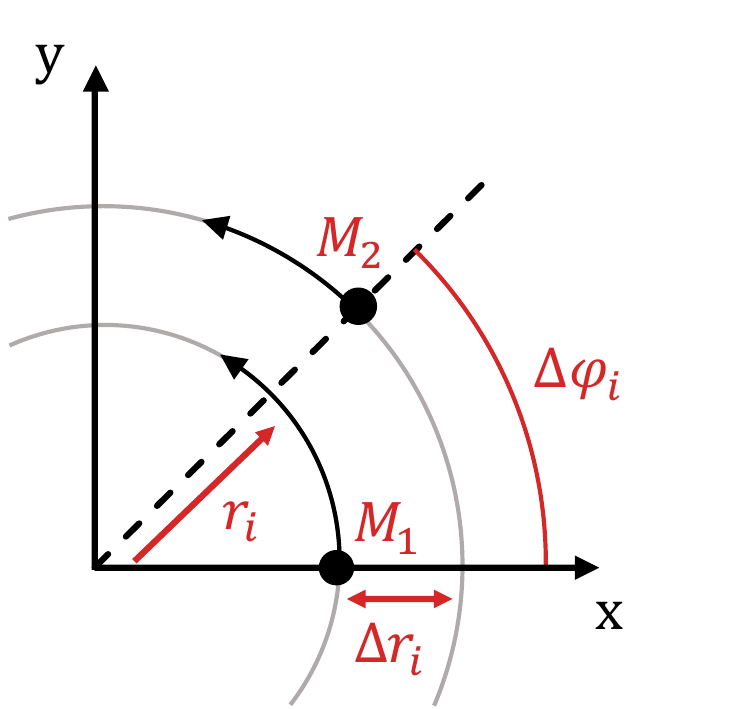}
    \caption{Parameters of an initial configuration, $r_{\mathrm{i}}$ the initial radius of the inner BH, $\Delta r_{\mathrm{i}}$ the radial separation between BHs, $\Delta \varphi_i$ the phase difference and $M_1, M_2$ the two masses.}
    \label{fig-schemaPara}
\end{figure}

Our aim is to probe the phase space of the initial conditions to determine the percentage of cases in which the BHs produce a Jacobi capture. A priori, we have 14 parameters to describe an initial configuration: 6 phase space coordinates per BH ($r, v_r, \theta, v_\theta, \varphi, v_{\varphi}$) and the mass of each object ($M_1$ and $M_2$). We decide to simplify the problem by considering the 2D case of two BHs orbiting in the same plane, which reduces our number of parameters to 10. We place the BHs on circular orbits in the $\theta=\pi/2$ plane which fixes $v_r$ and $v_{\varphi}$ once the position is known. Finally, we do not need each BH's phase ($\varphi_1$ and $\varphi_2$) but only the phase difference $\Delta \varphi$ because of cylindrical symmetry. We thus end up with five initial parameters: $r_1, r_2, \Delta \varphi_{\rm i}, M_1$ and $M_2$. However, it is more convenient to use $r_{\mathrm{i}}, \Delta r_{\mathrm{i}}, \Delta \varphi_{\rm i}, M_1$ and $M_2$ with $r_{\mathrm{i}}$ the initial radius of the inner BH and $\Delta r_{\mathrm{i}}$ the initial radial separation between the two BHs (these parameters are shown in Fig. \ref{fig-schemaPara}). Without loss of generality we impose that $M_1$ is always the inner BH (this means that $\Delta r_{\mathrm{i}}$ is always positive). We quantify the effect of the initial parameters on Jacobi capture and the percentage of cases in which such an event occurs. Parameter ranges are summarized in Table \ref{tab-5para}.
\begin{table}
\caption{Parameter ranges for the five-parameter case. For a definition of the parameters in use, see Fig.~\ref{fig-schemaPara}}             
\label{tab-5para}      
\centering                          
\begin{tabular}{c c c c c}        
\hline\hline                 
\multicolumn{1}{l}{} & \begin{tabular}[c]{@{}c@{}}Minimum\\ value\end{tabular} & \begin{tabular}[c]{@{}c@{}}Maximum\\ value\end{tabular} & \begin{tabular}[c]{@{}c@{}}Number\\ of values\end{tabular} & Spacing \\
\hline                        
   $M_1 / M_\odot$ & $10^3$ & $10^6$    & $8$  & Log \\      
   $M_2 / M_\odot$ & $10^3$ & $10^6$    & $8$  & Log \\
   $r_{\mathrm{i}} / \mathrm{kpc}$ & $0.1$  & $6$       & $14$ & Log \\
   $\Delta r_{\mathrm{i}} / \mathrm{kpc}$ & $0.08$ & $1.2$     & $30$ & Log  \\
   $\Delta \varphi_i$ & $0$    & $2\pi$ & $9$  & Lin  \\ 
\hline                                   
\end{tabular}
\end{table}

\subsection{Core stalling}
\label{coreStalling}

The general five-parameter case helps us emphasize the influence of masses and kinematics on Jacobi captures. Yet, to derive a probability of capture, we must delve deeper into a physically motivated parameter space with a reduced number of degrees of freedoms. To achieve this, we conduct a second simulation, assuming that BHs are off-center at their core stalling radius. This fixes the radii once the masses are defined, effectively reducing the number of free parameters to three. The stalling radius (noted $r_{\mathrm{cs}}$) is directly related to the mass of the BH through \citep[see][]{Kaur2018}
\begin{equation}
    \Omega_\mathrm{p}(r_{\mathrm{cs}})=\Omega_\mathrm{b} \quad \hbox{with}\quad \Omega_\mathrm{p}(r)=\sqrt{G \frac{M_\mathrm{b}(<r)+ M_\mathrm{p}}{r^3}},
    \label{eq-stallingRadius}
\end{equation}
where $\Omega_\mathrm{p}$ is the BH's orbital frequency, $\Omega_\mathrm{b}$ the orbital frequency of stars at the very center of the cored potential and $M_\mathrm{b}(<r)$ the enclosed mass of our background density. 
Numerical integration of Eq.~\ref{eq-stallingRadius} leads to a near-power solution which we find is accurately recovered from 
\begin{equation} \frac{r_\mathrm{cs}}{\textrm{kpc}} \simeq A \left(\frac{M}{M_\odot}\right)^{1/4} \label{eq-rcs}\end{equation} with $A \simeq 1 / 80$ a dimension-less constant.
We note that $r_{\mathrm{cs}}$ increases with the mass of the perturber, which is in agreement with \cite{Read2006}.
We end up with only three parameters: $M_1, M_2 \hbox{ and } \Delta \varphi_{i}$ (parameter ranges used are summarized in Table \ref{tab-3para}).
\begin{table}
\caption{Parameter ranges for the three-parameter case.}             
\label{tab-3para}      
\centering                          
\begin{tabular}{c c c c c}        
\hline\hline                 
\multicolumn{1}{l}{} & \begin{tabular}[c]{@{}c@{}}Minimum\\ value\end{tabular} & \begin{tabular}[c]{@{}c@{}}Maximum\\ value\end{tabular} & \begin{tabular}[c]{@{}c@{}}Number\\ of values\end{tabular} & Spacing \\
\hline                        
   $M_1 / M_\odot$ & $10^3$        & $10^6$ & $20$   & Log \\      
   $M_2 / M_\odot$ & $10^3$        & $10^6$ & $20$   & Log \\ 
   $\Delta \varphi_i$ & $2\pi / 1000$ & $2\pi$ & $400$  & Lin  \\ 
\hline                                   
\end{tabular}
\end{table}
We can impose $M_2 \geq M_1$ without loss of generality, ensuring that $M_1$ consistently represents the inner BH, as in the previous section. Not imposing this condition would mean simulating all possible cases twice, merely swapping the BH labels. This approach would yield mass-symmetric outcomes, offering limited insights into the impact of BH positioning on captures. Thus, we consider a total of $84\ 000$ initial configurations.

\subsection{Black hole radius of influence}
\label{sec-RH}

The volume of influence, also known as the Hill volume, defines a spatial domain surrounding the BH where its gravitational influence dominates. Traditionally, this volume is approximated as a sphere, characterized by the Hill radius (or radius of influence). In the classical restricted three-body problem, this radius is derived by determining the distance between the body of interest and the Lagrange point L1, situated along the line connecting the two massive bodies, between them. Under the distant-tide approximation, assuming a Hill radius much smaller than the distance from the body to the center and a significant mass hierarchy, an analytical expression for the Hill radius emerges: $R_H = (m/3M)^{1/3}R$, where $m$ represents the mass of the body of interest, $R$ its distance from the center, and $M$ the mass of the central body (\BT08 Sect. 8.3.1, Eq. 8.91). \\

To address the extended galactic case, using the previous analytical expression for the Hill radius, which involves replacing $M$ by $M_\mathrm{b}(<R)$ the enclosed mass of our background density at the radius of the BH, is not desirable. Complications arise due to several reasons: $(i)$ our investigation deviates from previous assumptions, particularly when we examine the dynamics of BHs in close proximity to the center ($R_H \sim R)$, where the galaxy's integrated mass is comparable in order to that of the BH ($M_\mathrm{b}(<R) \sim M_\mathrm{BH}$); $(ii)$ in density profiles featuring a central core, distinct Lagrange point structures emerge, differing from those observed in the Keplerian case \citep{Banik2022}; $(iii)$ the density profile employed lacks an analytical formula for the potential. For these reasons, we have conducted numerical computations to determine the Lagrange points and the radius of influence.\\

\begin{figure*}
    \centering
    \includegraphics[width=1.5\columnwidth]{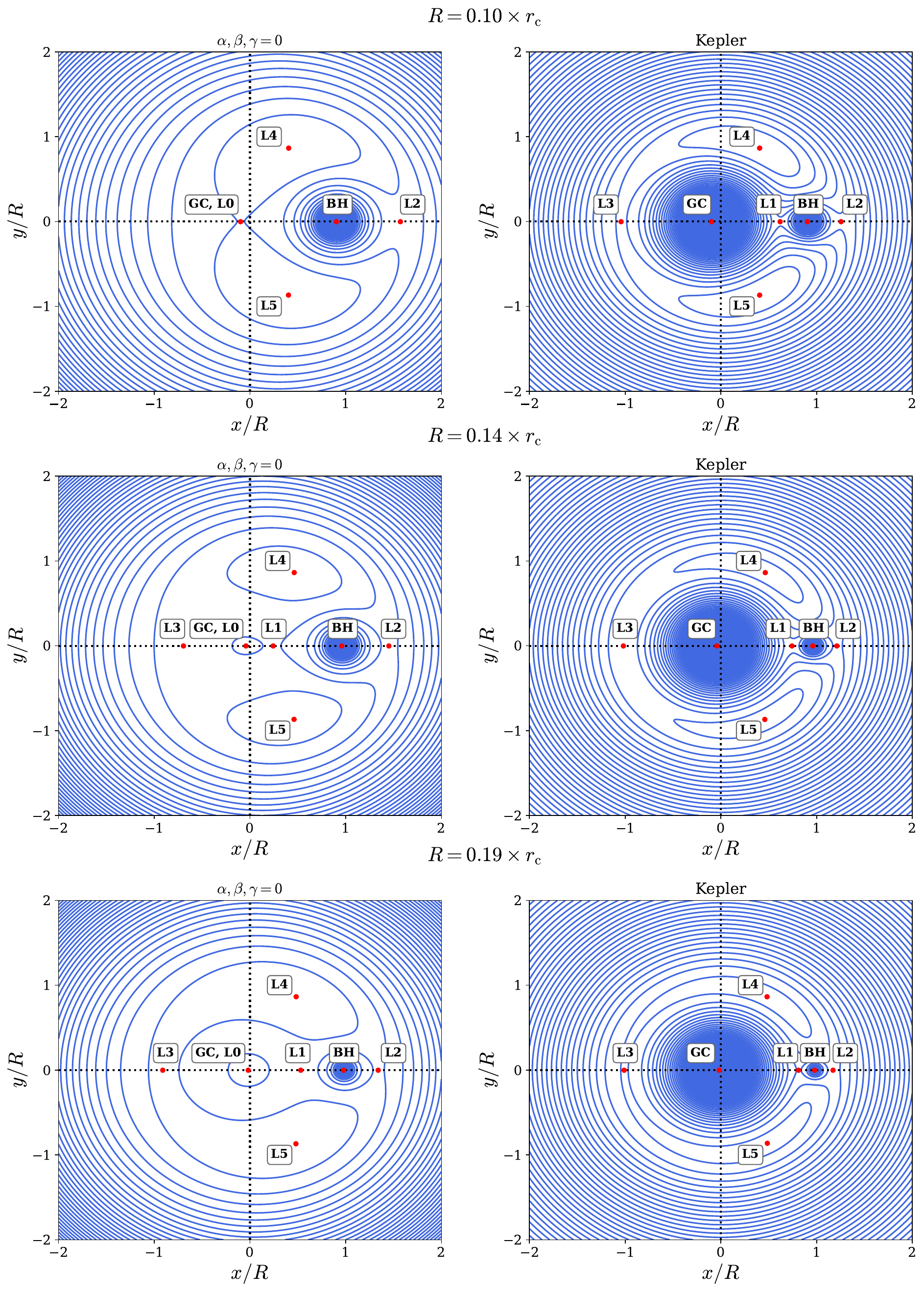}
    \caption{Effective potential of the galaxy + BH system plotted in the reference frame rotating at the angular frequency of the BH orbit in the $(x,y)$ plane. The coordinates $(x,y)=(0,0)$ denote the center of mass, calculated for the sum of the BH mass and the mass of the host galaxy integrated up to the BH orbit. Hence the galactic center is shifted significantly from $(0,0)$ in some cases. The Lagrange points, along with the positions of the black hole (denoted BH) and galactic center (denoted GC), are highlighted in red. $R = f \times r_{\mathrm{c}}$ is the orbital radius of the BH, where $f$ is a dimensionless factor and $r_{\mathrm{c}}$ represents the core radius of the potential, defined by its inflection point. We observe distinct differences in the topology of the effective potential, as well as in the positioning and number of Lagrange points, between the cored galactic potential (on the left) and the Keplerian case (on the right) where we approximate the galaxy as a point mass equal to the enclosed galactic mass up to the BH orbit.}
    \label{fig-phiEff}
\end{figure*}

Figure \ref{fig-phiEff} shows the effective potential of the galaxy + BH system, plotted in the reference frame rotating with the BH. Each row in Fig. \ref{fig-phiEff} corresponds to a distinct position of the BH within the core. The left-hand column is computed for the cored galactic potential, while the right-hand column corresponds to the Keplerian case, where we approximate the galaxy as a point mass equal to the enclosed galactic mass up to the BH orbit. Near the galactic center, the topology of the effective potential in the ($\alpha, \beta, \gamma$) case differs from the Keplerian case. As the BH approaches the center and reaches a critical radius $R_\mathrm{crit}$, a bifurcation occurs and the Lagrange points L1, L3 and L0 merge into a single point. In the last two rows of Fig. \ref{fig-phiEff}, we observe the scenario where the BH's radius exceeds this critical value, resulting in the same number of Lagrange points as in the Keplerian case, albeit at different positions. Conversely, the first row illustrates the case where the BH's radius is smaller than the critical value, leading to the merging of Lagrange points L1, L3, and L0\footnote{A detailed investigation is warranted to confirm whether this convergence indeed represents a single point or falls below our resolution limit. This aspect will be addressed in future research endeavors.}. As the BH's orbital radius increases, the galactic potential gradually approaches the Keplerian model.\\

\begin{figure}
    \centering
    \includegraphics[width=0.95\columnwidth]{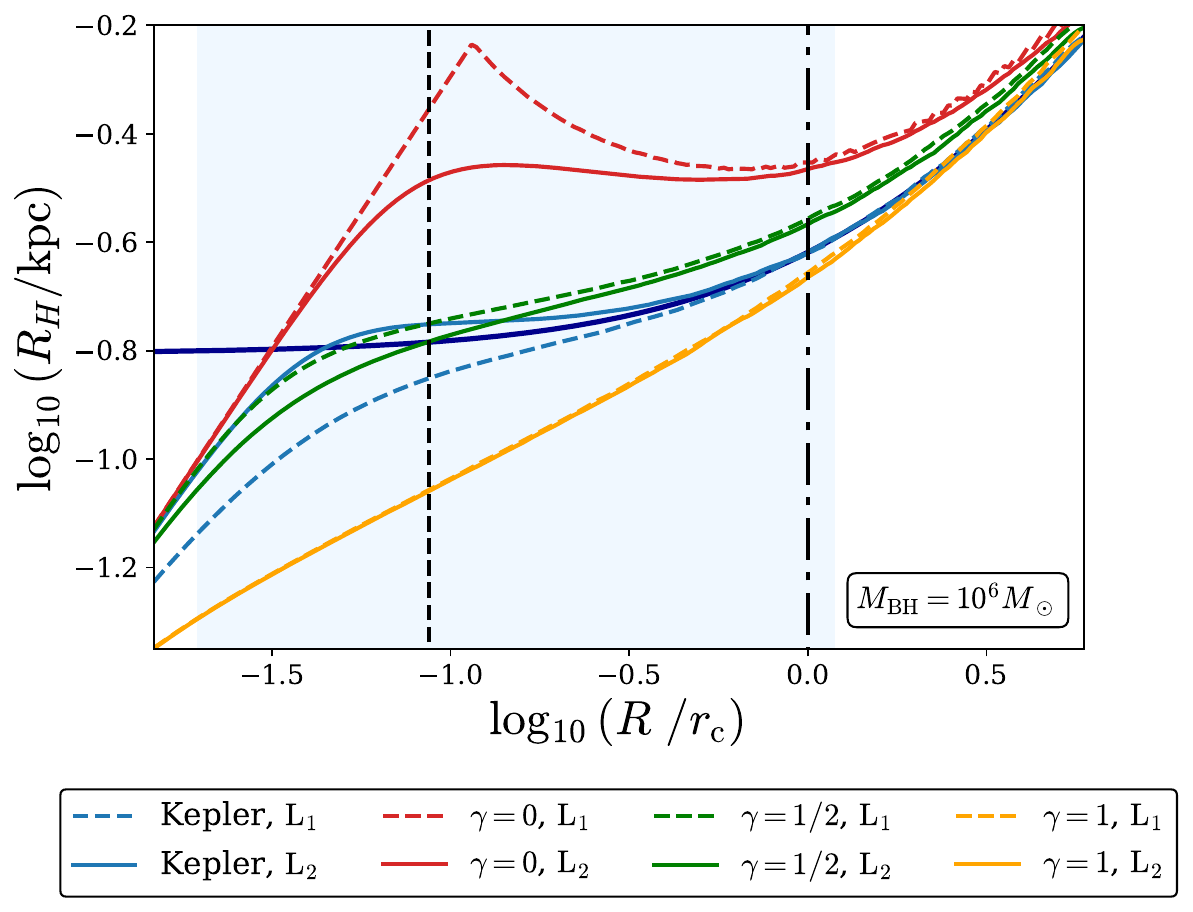}
    \caption{Hill radius ($R_\mathrm{H}$) as a function of the black hole (BH) orbital radius $R$ in units of the core radius for a mass $M_\mathrm{BH}=10^6 M_\odot$. The Hill radius is computed as the BH-L1 distance (dashed lines) and the BH-L2 length (solid lines). Four background potentials were considered : a Keplerian potential (shown as light-blue lines); and three $\alpha, \beta, \gamma$ profiles with $\alpha=1$, $\beta=3$ and inner slopes $\gamma=0$ (red), $1/2$ (green) and $1$ (yellow) respectively. The thick dark blue solid line is for the distant-tide approximation, which goes to a constant value as $R \rightarrow 0$. On the figure, the shaded area denotes the scope of this study. The  core stalling radius and galactic core radius are indicated by vertical lines. The separation of the dashed and solid lines indicates an asymmetry between L1 and L2 on either side of the black hole when entering the core region. Note that for the case of an NFW profile ($\gamma=1$ , yellow) the L1 and L2 solutions very nearly overlap inside the core. The discontinuity of the dashed red line shows a bifurcation of the Lagrange points in the core profile. It is striking that the Keplerian potential remains close to the distant-tide approximation well inside the core of the galaxy, before diverging significantly at $\log \ R/r_\mathrm{c} \lesssim -1.3$. By comparison, extended profiles may show differences with this approximation exceeding a factor 2 anywhere inside the core for $\gamma=0$ (red).}
    \label{fig-RH}
\end{figure}

Figure \ref{fig-RH} shows the Hill radius as a function of the BH orbital radius. It is computed in two ways, by taking the distance between the BH and the L1 Lagrange point and the distance between the BH and L2. Four background potentials were considered : a Keplerian potential and three ($\alpha, \beta, \gamma$) profiles with different inner density slopes.\\

First, we observe that the distance between the BH and L1 matches that between the BH and L2 around and beyond the core radius. However, this symmetry does not persist within the core. Here, the distant-tide approximation, which enforces precise L1/L2 symmetry by eliminating asymmetry terms in the calculations, breaks down. The exception is the NFW profile ($\gamma = 1$), where symmetry holds across all radii.\\

In our study profile with a core ($\gamma = 0$), we discern a discontinuity in the BH-L1 distance at a specific radius, indicative of the earlier mentioned bifurcation. Beyond this critical radius, the L1 Lagrange point lies between the BH and the galactic center. However, as the BH's orbital radius diminishes, the L1 point converges toward the galactic center. Below the critical radius, the distance between the BH and the galactic center is measured, accounting for the linearity of the curve.\\

The Keplerian profile closely aligns with the Hill radius predicted by the distant-tide approximation formula, maintaining this consistency even within the core despite L1/L2 asymmetry. It is only in very close proximity to the center that significant divergence occurs. In contrast, the extended profiles ($\alpha, \beta, \gamma$) exhibit clear discrepancies with the distant-tide approximation, particularly within the core, notably for inner slopes $\gamma=0$ and $\gamma=1$. Within our study profile ($\gamma=0$), we see a difference of a factor of 2 between the distant-tide approximation's analytical formula and the precise computation. This discrepancy, coupled with the distinct shape of the Hill radius curve compared to the Keplerian case, suggests that the galaxy cannot be adequately treated as a radius-dependent point mass.
Moreover, the disparities among the three extended density profiles highlight the significant dependence of the Hill radius in our study area on the inner density slope. Consequently, we anticipate an equally pronounced sensitivity of the capture process to this parameter.\\

Figure \ref{fig-phiEff} and \ref{fig-RH} show that the dynamics and likelihood of Jacobi captures are significantly influenced by the galactic context. Firstly, the zone of influence of BHs is different from the Keplerian case, and secondly, the altered topology of the effective potential gives rise to novel types of BH orbits, including Pacman orbits (see Fig. \ref{fig-pacman}), which are pivotal for core stalling \citep{Banik2022}.\\

\subsection{Capture criterion}

We aim to contextualize Jacobi captures within an observable framework to evaluate the likelihood of forming and detecting off-center BH binaries. To achieve this, we conduct simulations spanning a timeframe of $14$ Gyr. Consequently, some captures may be truncated by halting the simulation, while others that are theoretically feasible may not have sufficient time to occur. For instance, when the orbital radii of BHs are extremely close, the time required to catch-up their phase difference may exceed $14$ Gyr. Likewise, in order to deduce a probability from initial conditions, we incorporate conditions that lead to capture at $t=0$ Gyr. These cases are indeed conditions allowing BH binding off-center, and due to the ergodic nature of this study, any stretch of $14$ Gyr should be equivalent to another. Put differently, a time shift in the simulation range would interchange initial and dynamical captures (for instance, initiating the simulation later would render some dynamical captures as initial and vice versa). Lastly, although we simulate $14$ Gyr of evolution during which an initial condition can result in several captures, we solely focus on the first, as higher-order effects can significantly influence the sequence of events in reality (such as dynamical friction and post-Newtonian corrections).\\

To determine a capture, two conditions must be met. First, the binding energy between the BHs must be negative. This is given by
\begin{equation}
    E_{\mathrm{b}} = \frac{1}{2}\mu \ v_{\mathrm{rel}}^2 - \frac{GM_1M_2}{d_{\mathrm{rel}}},
    \label{eq-bindingEnergy}
\end{equation}
with $\mu = M_1M_2/(M_1+M_2)$ the reduced mass, $G$ the gravitational constant, $d_{\mathrm{rel}}$ the relative distance between the two BHs and $v_{\mathrm{rel}}$ their relative velocity. Secondly, during the period of negative binding energy, the distance between the BHs must fall below the binary Hill radius. This radius is defined as the average distance between BH-L1 (or BH-L0 if L1 no longer exists) and BH-L2 for a total mass $M_\mathrm{tot}=M_1+M_2$ located at $\Tilde{r} = (M_1\cdot r_1 + M_2\cdot r_2) / (M_1+M_2)$. This additional condition prevents mislabeling due to coincidental alignment of velocity vectors, which could lead to a negative energy reading. It is important to bear in mind that classical two-body dynamics do not allow for gravitational capture, as BHs get accelerated upon approach due to energy conservation. However, in the three-body system, the presence of a third body enables the binding energy to dip below zero and later return to positive values, allowing for unbinding \citep{PetitHenon86, renaud2011, Penarrubia2023}. The time spent in this sub-zero state, during which the BHs can revolve around each other a few times, is challenging to predict due to the chaotic nature of the three-body problem, occurring over variable timescales. However, dissipative forces like dynamical friction, three-body scattering or post-Newtonian corrections can enhance temporary captures when considered \citep{Samsing2018, Li2022, Boekholt2023, Rowan2023, DeLaurentiis2023}. Consequently, since our focus in this study is on Jacobi captures, our aim is to identify sub-zero crossings independently of their duration.\\

Although flybys with a single close encounter are traditionally distinguished from captures with more than one close encounter, we opt to label both scenarios as captures in this article: ($i$) unlike other distance-based criteria \citep{Boekholt2023}, Eq. \ref{eq-bindingEnergy} is more stringent as it incorporates a velocity condition for BHs. This refinement ensures that the surviving flybys consist of BHs close in energy and with mean eccentricities below 1 (see Fig. \ref{fig-ecc}). Consequently, we anticipate that these flybys may be more easily stabilized in the future (for a comparison with another, less restrictive criterion yielding numerous hyperbolic flybys, refer to Section \ref{sec-crit}); ($ii$) it is conceivable that even a single close encounter could lead to eventual stabilization, especially when dissipative forces, such as those arising from the post-Newtonian corrections, play a significant role at the pericenter of eccentric binaries (cf. Fig. \ref{fig-ecc}). The question of the conditions necessary for stabilization remains a subject for future investigation.\\

\begin{figure*}
    \centering
    \includegraphics[width=1.7\columnwidth]{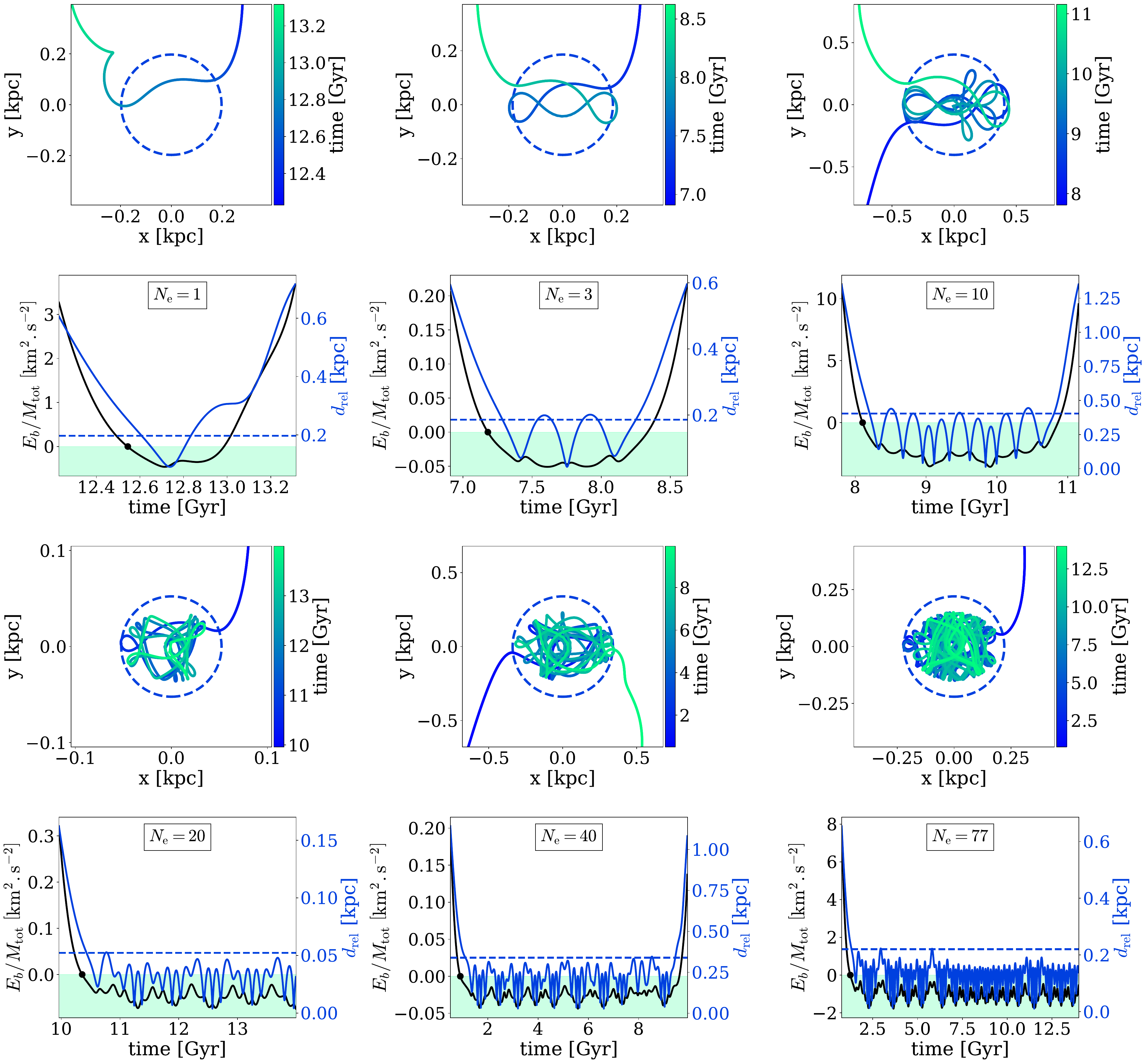}
    \caption{Examples of Jacobi captures with different numbers of close encounters between BHs. Top row: relative coordinates between BHs during Jacobi captures. The reference frame is centered on the most massive BH and rotates with it. The binary Hill radius is marked by the dotted circle. Bottom row: the black line represents the binding energy ($E_\mathrm{b}$) per unit mass of the binary ($M_\mathrm{tot}=M_1+M_2$), with negative values shown in the colored area. The black marker indicates the point when the energy becomes negative. The blue curve represents the relative distance ($d_\mathrm{rel}$) between the BHs. The binary Hill radius is indicated by a dashed blue line. The number of close encounters, $N_\mathrm{e}$, is given in the box. To contextualize captures within observational framework, we extend our simulations over a period of $14$ Gyr. Consequently, we observe instances where captures are halted prematurely, as illustrated by cases with $N_\mathrm{e}=20$ and $N_\mathrm{e}=77$. Additionally, we retain flybys due to our energy-based criterion, which is more stringent compared to distance-based criteria. This ensures that surviving flybys comprise black holes with similar energy levels, potentially facilitating stabilization through dissipative forces.}
    \label{fig-trajRH}
\end{figure*}

Figure \ref{fig-trajRH} illustrates six instances of Jacobi captures characterized by varying numbers of close encounters. The top row displays the relative coordinates between the BHs during these captures, with the reference frame centered on the most massive BH and rotating with it. The binary Hill radius is depicted by the dotted circle. The bottom row shows the binding energy and the relative distance between the BHs.

\subsection{Chaos}
\label{sec-chaos}

\begin{figure*}[htbp]
    \begin{minipage}[c]{0.28\linewidth}
        \centering
        \includegraphics[width=\linewidth]{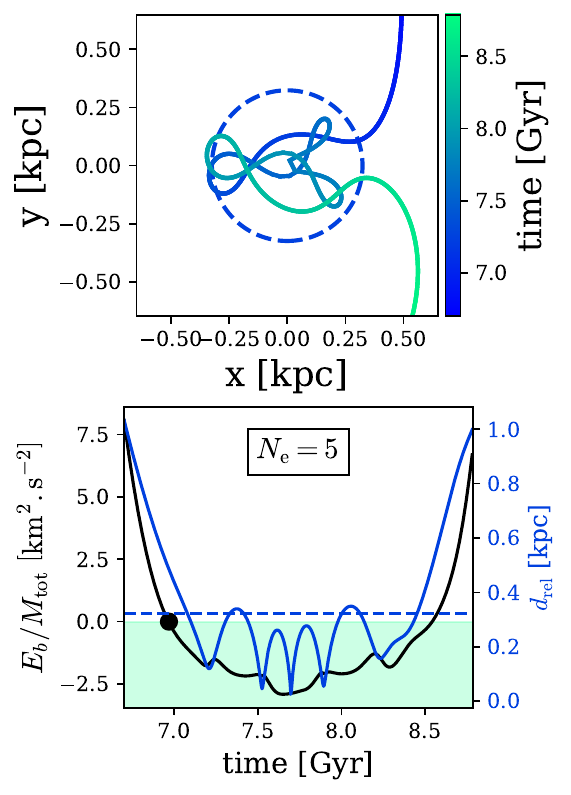}
    \end{minipage}
    \hfill
    \begin{minipage}[c]{0.70\linewidth} 
        \centering
        \includegraphics[width=\linewidth]{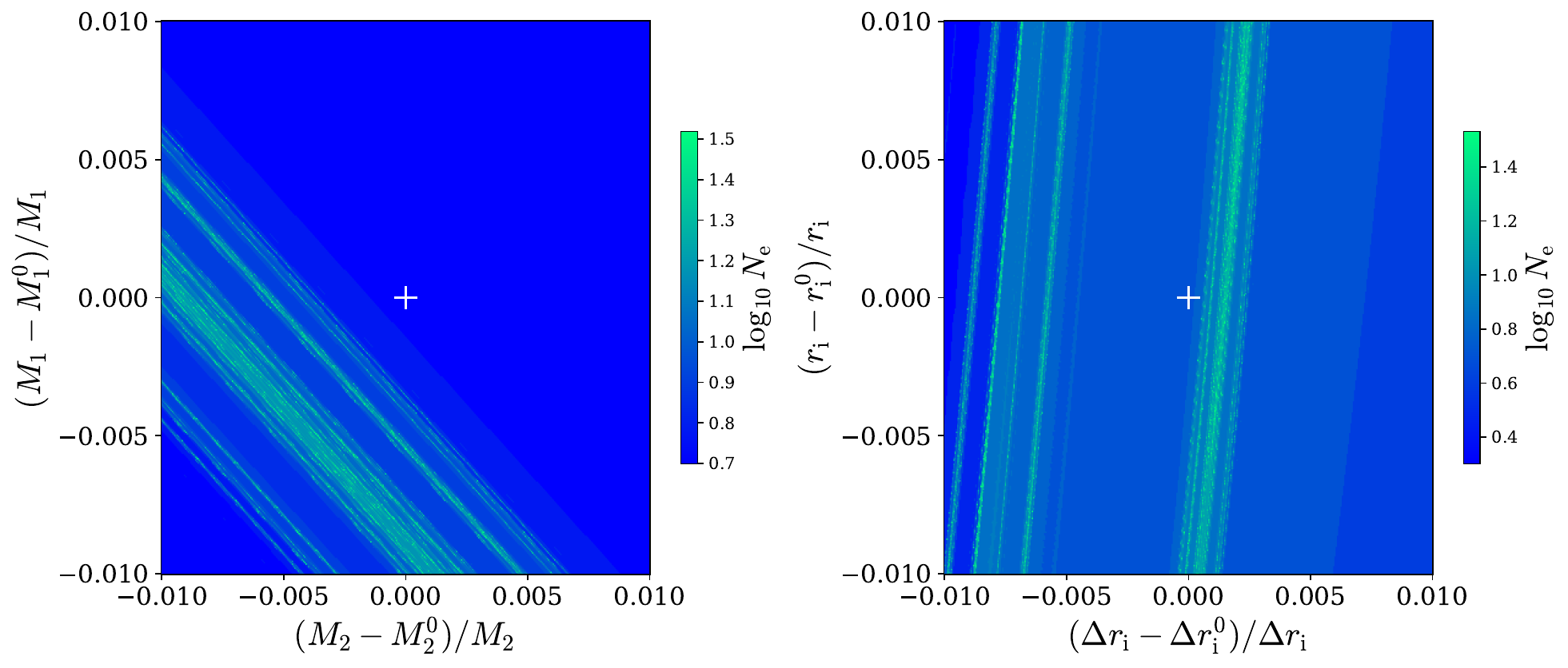}
    \end{minipage}
    \caption{Influence of the initial conditions on the number of close encounters ($N_\mathrm{e}$). Left panel: Reference Jacobi capture ($M_1^0=3.7276\times 10^5\ M_\odot$, $M_2^0=3.7276\times 10^5\ M_\odot$, $r_\mathrm{i}^0=6.000 \ \mathrm{kpc}$, $\Delta r_\mathrm{i}^0=0.6242 \ \mathrm{kpc}$, $\Delta \varphi_\mathrm{i}^0=3.4907$, with significant figures up to the resolution of the central and right panels). The top row shows the relative coordinates between the BHs during the capture. The reference frame is centered on the most massive black hole and rotates with it. The binary Hill radius is indicated by the dotted circle. In the bottom row, the black line represents the binding energy ($E_\mathrm{b}$) per unit mass of the binary ($M_\mathrm{tot}=M_1+M_2$), with negative values depicted in the colored area. The black marker indicates the point at which the energy becomes negative. The blue curve represents the relative distance ($d_\mathrm{rel}$) between the BHs. The binary Hill radius is shown by a dashed blue line. The reference capture involves 5 close encounters between the black holes. Central panel: $N_\mathrm{e}$ variations as a function of initial conditions ($M_1, M_2$). The ($M_1, M_2$) values are distributed plus or minus $1\%$ around the initial condition of the reference capture, denoted as ($M_1^0, M_2^0$). Right panel: $N_\mathrm{e}$ variations as a function of initial conditions ($r_\mathrm{i}, \Delta r_\mathrm{i}$). The ($r_\mathrm{i}, \Delta r_\mathrm{i}$) values are distributed plus or minus $1\%$ around the initial condition of the reference capture, denoted as ($r_\mathrm{i}^0, \Delta r_\mathrm{i}^0$). The reference capture is located at ($0,0$) in each panel. The strong variations of $N_\mathrm{e}$ with initial conditions is suggestive of chaotic motion.}
    \label{fig-chaos}
\end{figure*}

To underscore the chaotic behavior of Jacobi captures in this context, we choose a specific capture as a point of reference and examine how small changes in initial conditions affect the number of close encounters between the BHs. In the left-hand panel of Fig. \ref{fig-chaos}, this reference capture is shown, characterized by 5 close encounters ($N_\mathrm{e}$) between the BHs. We set the initial conditions ($M_1, M_2, r_\mathrm{i}, \Delta r_\mathrm{i}, \Delta \varphi_\mathrm{i}$) for this capture and conducted two refined simulations: one varying ($M_1, M_2$) more finely and another varying the radial parameters ($r_\mathrm{i}, \Delta r_\mathrm{i}$). For both sets of parameters, we sampled 400 linearly distributed values, varying them by plus or minus $1\%$ relative to the original initial conditions. The central and right-hand panels of Fig. \ref{fig-chaos} present the number of close encounters between the BHs for each case. In these regions of phase space, the number of close encounters fluctuates between $5$ to $33$ for the central panel and $2$ to $34$ for the right-hand panel. The upper limits are due to the fixed integration time, truncating the captures beyond 14 Gyr. The significant irregularities in the count of close encounters when both pairs of parameters are varied illustrate the high sensitivity to initial conditions. On the central panel, the patterns of irregularity align along lines where $M_1 + M_2 \sim C$, with $C$ being a constant. This occurs because at a larger radius (here, the internal BH is $6$ kpc from the galactic center), dynamics is governed by the binary Hill radius, which depends on the total mass.

\section{Results}
\label{results}

\subsection{Bound black holes in the general case}
\label{generalCase}

For the general case where we made no assumption regarding the cause of the offset from the center, we integrated $8\times 8\times 14\times 30\times 9 = 241\,920$ initial configurations (cf. Table \ref{tab-5para}) over 14 Gyr. We find a capture in $7.1\%$ of cases, of which $26.2\%$ are flybys. These captures can be further categorized into two distinct types: those formed at the beginning of the simulations (t=0 Gyr, hereafter referred to as initial captures), comprising $66.3\%$ of the total captures, and those emerging during the integration at later times (t>0 Gyr, hereafter referred to as dynamical captures), making up the remaining $33.7\%$ of the captures. It is essential to bear in mind that in this section our focus is on the initial conditions that initiate the capture process i.e., the transition from two separate two-body problems to a three-body problem. Consequently, the results presented here are not subject to chaos. We examine the initiation of the process rather than the subsequent three-body interaction (such as capture duration or number of close encounters) which typically exhibit chaotic behavior (cf. Section \ref{sec-chaos}).\\

\begin{figure*}[h!]
	\begin{minipage}{0.73\hsize}
		\includegraphics[width=1.\columnwidth]{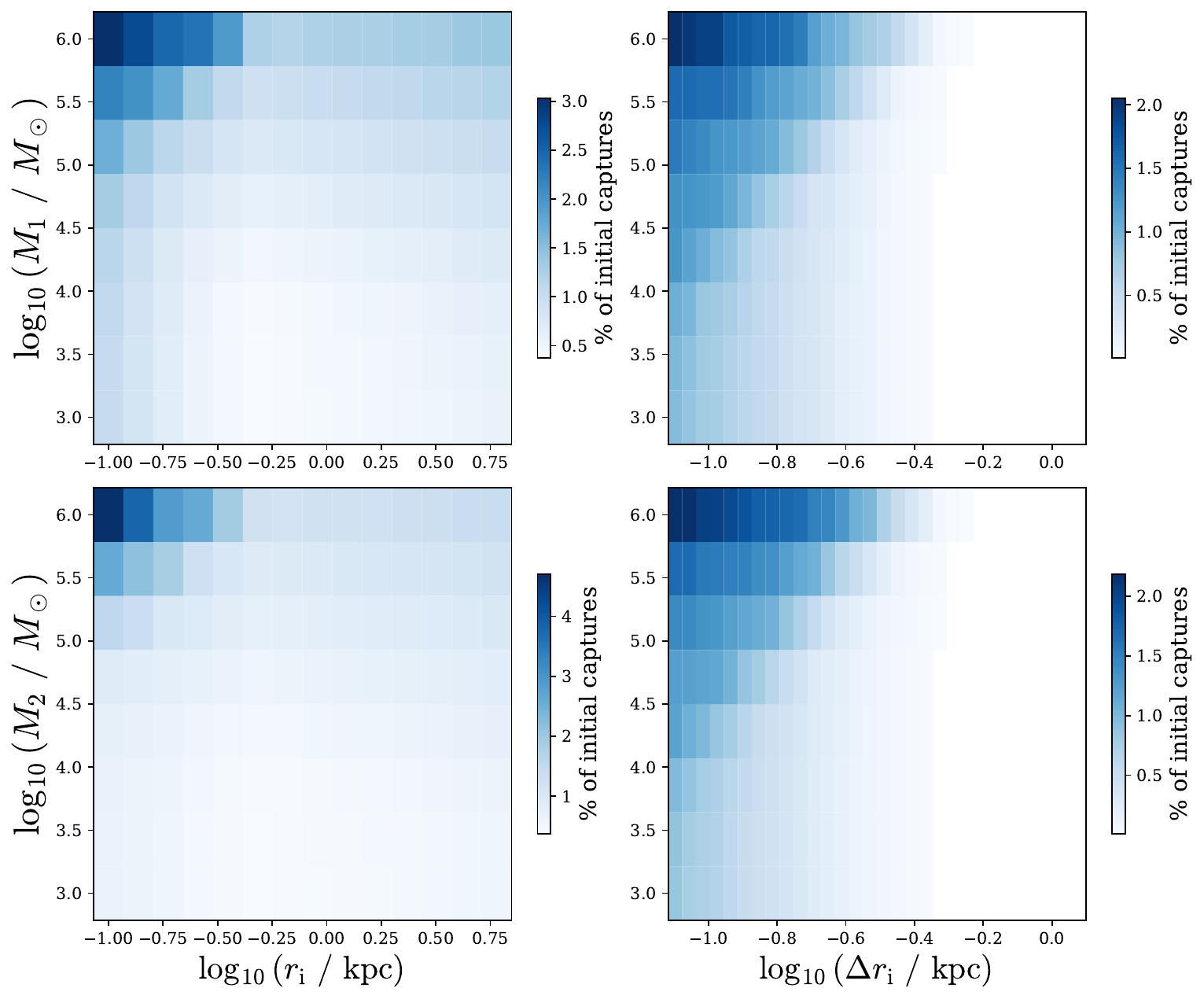}
	\end{minipage}
 \hfill
	\begin{minipage}{0.02\hsize}
	
	\end{minipage}
	\begin{minipage}{0.23\hsize}
		\caption{Percentage of initial captures as a function of different parameter pairs, marginalising each time over the three others. Top-left panel: as a function of the mass of the inner BH ($M_1$) and its initial radius ($r_\mathrm{i}$); bottom-left panel: as a function of the mass of the outer BH ($M_2$) and $r_\mathrm{i}$; top-right panel: as a function of $M_1$ and the initial radial separation between the BHs ($\Delta r_\mathrm{i}$); bottom-right panel: as a function of $M_2$ and $\Delta r_\mathrm{i}$.}
    \label{fig-iniCapture}
	\end{minipage}
    \end{figure*}

Figure \ref{fig-iniCapture} shows the percentage of initial captures in the space of different parameter pairs, marginalized over the others. The right-hand panels show this percentage in $(M_1,\Delta r_\mathrm{i})$ space (top) and in $(M_2,\Delta r_\mathrm{i})$ space (bottom). In both panels, the proportion of captures increases as the radial separation between the BHs decreases and their masses increase. These conditions elevate the potential term of the binding energy (see Eq. \ref{eq-bindingEnergy}), consequently reducing the binding energy of the binary. Beyond a certain threshold of radial separation, captures cease to occur. This limit does not precisely match the maximum binary Hill radius (for the parameter range explored in this study, $\log R_\mathrm{H}^\mathrm{max} = -0.35$), as it is possible that initially, the BHs are slightly beyond the sphere of influence radius but their relative velocity may be low enough to initiate the capture process. Nevertheless, the maximum radial separation allowing initial capture is rather close to $R_\mathrm{H}^\mathrm{max}$.\\

The top left panel of Fig. \ref{fig-iniCapture} shows the percentage of initial captures as a function of $(M_1,r_\mathrm{i})$, while the lower left panel shows it as a function of $(M_2,r_\mathrm{i})$. Regarding the $r_\mathrm{i}$ dependence, we observe a sharp increase in the percentage of capture at low radii ($\log r_\mathrm{i} / \mathrm{kpc} < -0.4$) for higher masses, which then stabilizes for higher radii. The constant part at higher radii is attributed to the nearly constant Hill radius within this range (refer to Fig. \ref{fig-RH} for the case of a $10^6 M_\odot$ BH), the potential term is therefore dominant in this zone (for further details, see Fig. \ref{fig-IniOnlyRH} and Appendix \ref{sec-crit}). At low radii, the probability increase cannot be explained by Hill's radius, which decreases in this zone. Instead, the kinetic component of the binding energy plays a significant role in these captures. In this low-radius zone, the gradient of circular velocities decreases due to the presence of the density core ($\gamma = 0$), leading to a reduction in the relative velocity between BHs and consequently lowering the binding energy. Another contributing factor is the marginalization over other parameters. At low radii, there are more values of $\Delta \varphi_\mathrm{i}$ that enable the BHs to remain within their sphere of influence, despite its decrease.\\

\begin{figure}
    \centering
    \includegraphics[width=0.8\columnwidth]{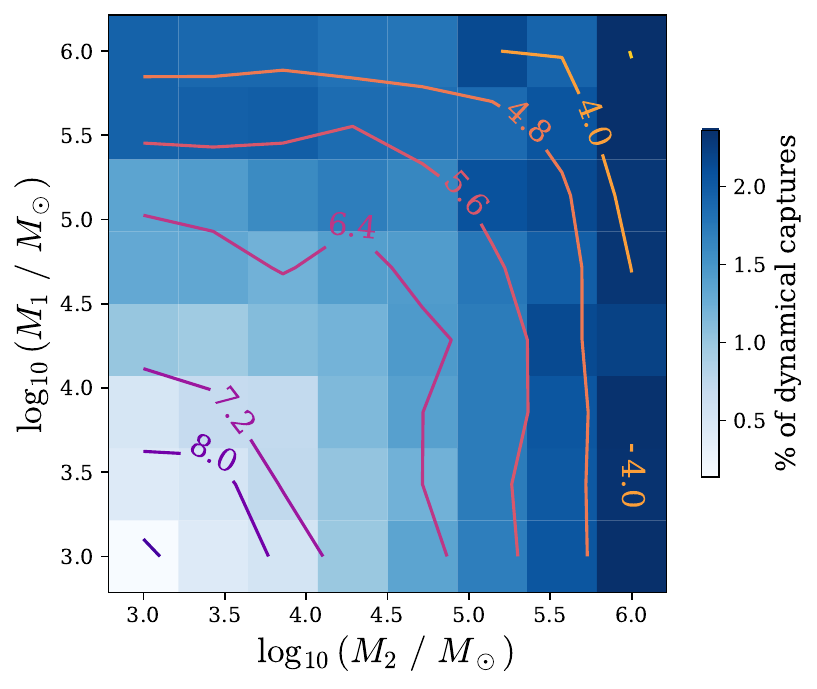}
    \caption{Percentage of dynamical captures as a function of inner ($M_1$) and outer ($M_2$) BH mass. The contour lines represent the average time it took for the capture to occur since the start of the simulation, measured in Gyr. The percentage of captures increases as the mass of the two BHs increases, and the mass of the outer BH is greater than the mass of the inner one ($M_2 > M_1$).}
    \label{fig-dynCaptureM1M2}
\end{figure}

Figure \ref{fig-dynCaptureM1M2} shows the percentage of dynamical captures (i.e., happening at t>0 Gyr) as a function of $(M_1, M_2)$. The percentage of captures increases as the mass of the two BHs increases, and we notice that the percentage is higher when the mass of the outer BH is greater than the mass of the inner one ($M_2 > M_1$). This is of particular interest within the context of core stalling, which predicts a larger stalling radius for more massive BHs (cf. Eq. \ref{eq-rcs}). This outcome can be attributed to the interaction between BHs, which imparts eccentricity to the initially circular orbits over time. This eccentricity gain depends on the BH mass: when one BH is much more massive, it remains nearly circular while the lighter one becomes eccentric. Consequently, if the more massive BH is positioned closer to the center, the lighter one will approach it more centrally, resulting in higher kinetic energy and a more challenging capture. On the contrary, when the more massive BH occupies the outer position, the lighter BH approaches it at a larger radius, resulting in reduced kinetic energy and thus facilitating the capture process. In Fig.~\ref{fig-dynCaptureM1M2}, we also overplotted contour lines representing the average time it took for the capture to occur since the start of the simulation. The average capture time shows an increase as the total BH mass diminishes.\\

\begin{figure*}[h!]
	\begin{minipage}{0.74\hsize}
		\includegraphics[width=1.\columnwidth]{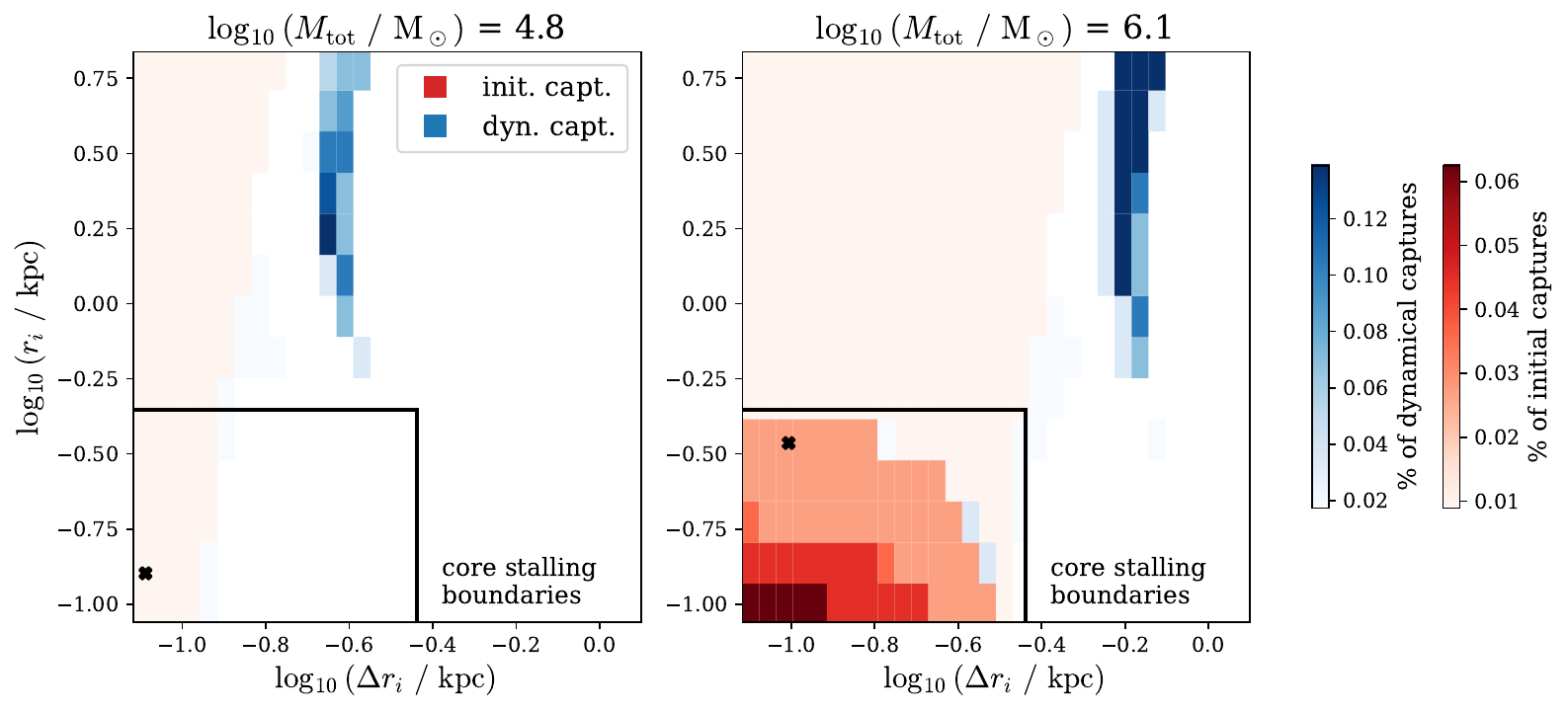}
	\end{minipage}
 \hfill
	\begin{minipage}{0.02\hsize}
	
	\end{minipage}
	\begin{minipage}{0.23\hsize}
		\caption{Percentage of captures as a function of initial inner BH radius ($r_{\mathrm{i}}$) and initial radial separation ($\Delta r_{\mathrm{i}}$) for two mass pairs, On the left panel, $\log (M_1/M_\odot) = 3.9$ and $\log (M_2/M_\odot) = 4.7$, on the right panel, $\log (M_1/M_\odot) = 5.6$ and $\log (M_2/M_\odot) = 6$. Red areas represent initial captures while blue areas dynamical captures. The black lines indicate the limits within which core stalling can position our BHs, given the mass range studied. The black crosses indicate where core stalling would position the BHs in each case.}
    \label{fig-frangesColor}
	\end{minipage}
    \end{figure*}

Figure \ref{fig-frangesColor} represents the percentage of captures as a function of $(r_{\mathrm{i}},\Delta r_{\mathrm{i}})$, when we fix the masses $M_1, M_2$ and marginalize over $\Delta \varphi_{\mathrm i}$ only. We show the results for two different pairs of masses, on the left-hand panel $\log (M_{\mathrm{tot}}/M_\odot) = 4. 8$ and on the right panel $\log (M_{\mathrm{tot}}/M_\odot) = 6.1$. Red areas represent initial captures and blue areas dynamical captures. Note the presence of blue bands that shift to the right ($\Delta r_{\mathrm{i}}$ larger) as the total mass increases. To the left of these bands ($\Delta r_{\mathrm{i}}$ smaller), we notice the presence of initial captures on a shallow red background. The plot can be divided into three regions: to the left, within, and to the right of the bands.\\

To the right of the band ($\Delta r_{\mathrm{i}}$ larger), there is no capture: the BHs are too distant initialy to reach a configuration where their gravitational attraction is comparable to the force generated by the external potential. To the left of the band ($\Delta r_{\mathrm{i}}$ smaller), captures occur at t=0 (in red). Indeed, two possible scenarios arise: $(i)$ captures form at t=0 if BHs have a sufficiently low initial phase shift $\Delta \varphi_\mathrm{i}$, $(ii)$ the initial phase shift is too large for an initial capture, and $\Delta r_{\mathrm{i}}$ is too small to allow phase catch-up within the simulation timeframe\footnote{When $\Delta r_{\mathrm{i}}$ is small, the BHs share similar orbital frequencies, so their phase difference evolves slowly.}. Also note the darker red area in the lower left of the right panel of Fig.~\ref{fig-frangesColor}. In this region the BHs are positioned very near the galactic center ($r_\mathrm{i}$ small) and are in close radial proximity to each other. Under these conditions, the shorter distances for the same angular separation allow for more flexibility in terms of the initial phase shifts that result in Jacobi capture. In other words, when $r_\mathrm{i}$ decreases larger initial phase shifts can lead to a capture, which explains the increased capture percentage.\\

\subsection{Probability of capture}
\label{CaptureProb}

The previous section provides valuable insights into the impact of various parameters on Jacobi captures. In this section, our objective is to reduce the number of free parameters by making two assumptions concerning the origin and decentering of BHs. This allows us to calculate a probability of capture.

\subsubsection{Bound black holes from the core stalling radii}
\label{KaurCase}

In order to reduce the number of free parameters, we first make the assumption that the BHs are off-center due to core stalling, which sets $r_{\mathrm{i}}$ and $\Delta r_{\mathrm{i}}$ (cf. Section \ref{coreStalling}), leaving only three free parameters (parameter ranges remaining unchanged). By employing this approach, we analyzed $84\,000$ simulations, $11\,712$ of which show a Jacobi capture, corresponding to $13.9\%$ of cases. There is an increase in the percentage of captures, meaning that core stalling  allows BHs to be placed on radii that favor capture.\\

\begin{figure*}[h!]
	\begin{minipage}{0.73\hsize}
		\includegraphics[width=1.\columnwidth]{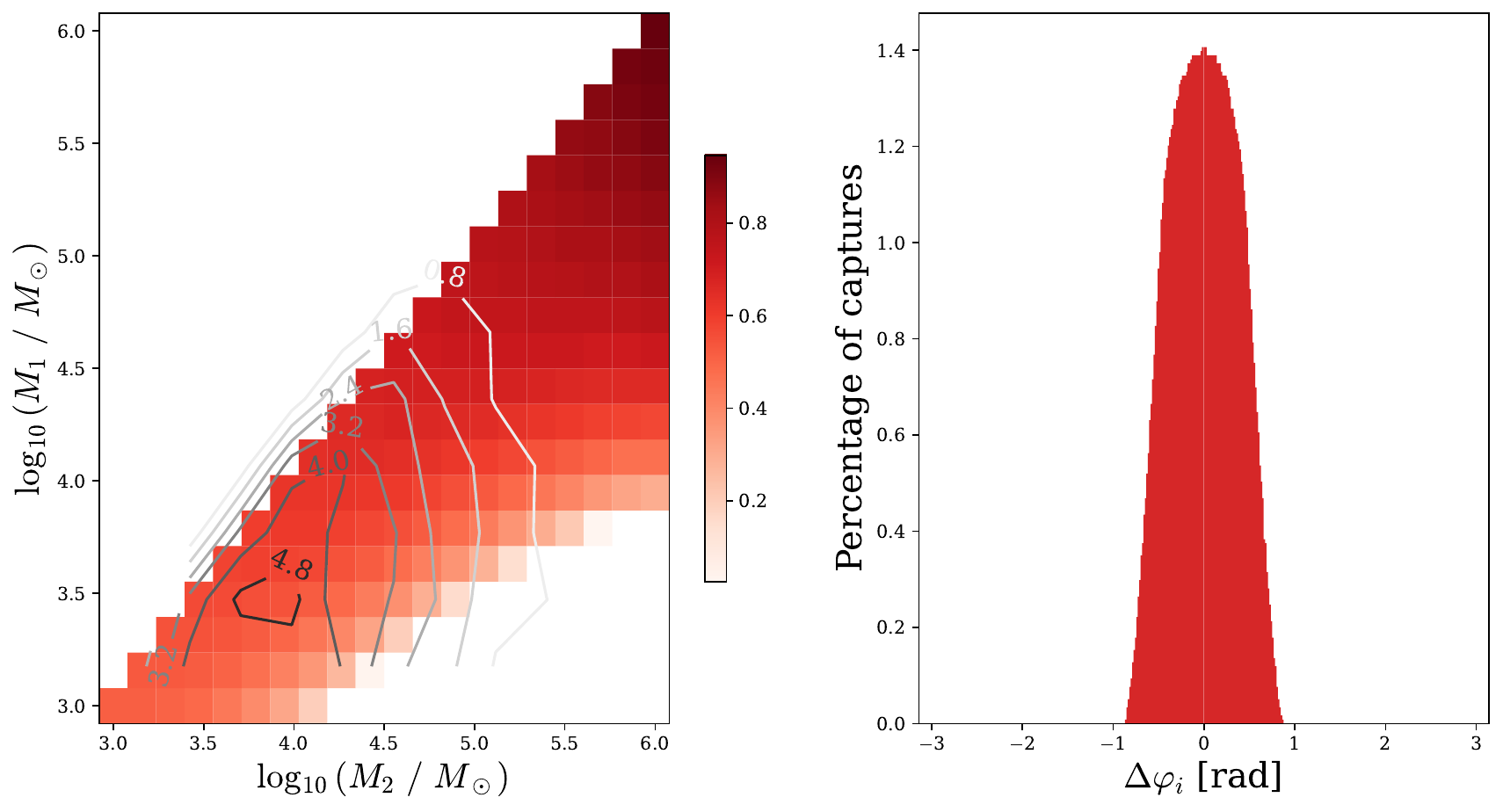}
	\end{minipage}
 \hfill
	\begin{minipage}{0.02\hsize}
	
	\end{minipage}
	\begin{minipage}{0.23\hsize}
		\caption{Percentage of captures in the three-parameter case. Left panel: as a function of inner ($M_1$) and outer ($M_2$) BH mass when positioned on their core stalling radius. The contour lines depict the probability (in percentage) of obtaining a mass pair (derived from a Monte Carlo sampling using SatGen, see Section \ref{satgen}). Right panel: as a function of the initial phase difference.}
    \label{fig-3D}
	\end{minipage}
    \end{figure*}

The left panel of Fig. \ref{fig-3D} shows the percentage of captures as a function of the mass of the inner ($M_1$) and outer ($M_2$) BH. The upper left region of the figure is left empty because of the condition $M_2 \geq M_1$, guaranteeing that BH2 consistently holds the outer position (cf. Section \ref{coreStalling}). The right-hand panel depicts the percentage of captures as a function of the initial phase difference $\Delta \varphi_i$. The left panel reveals a higher percentage of captures when both $M_1$ and $M_2$ exhibit higher values, and when the mass gap between BHs is modest. The right-hand panel illustrates that captures predominantly occur with small initial phase difference ($\Delta \varphi_i$ close to 0).\\

Core stalling positions the BHs on very similar radii, as can be seen in Fig. \ref{fig-frangesColor}, where the black lines indicate the boundaries within which core stalling operates for the studied mass range of $10^3-10^6\, M_\odot$. This region is in the regime of small radial separation discussed in Section \ref{generalCase}. In that region, the vast majority of captures ($95.4\%$) occur initially, when the BHs are sufficiently close in both initial radius and initial phase shift. Furthermore, to understand why there is a higher percentage of capture when masses are high, we can once again refer to Fig. \ref{fig-frangesColor}. In each panel, the black cross indicates the positioning of BHs by core stalling. Notably, in the right-hand panel with higher masses, this cross falls within a darker region. This zone shifts from left to right as the mass of the BHs increases. In the right panel of Fig. \ref{fig-3D}, captures are observed at low angular separations, as we simulate 14 Gyr of evolution to evaluate captures within a Hubble time. When the angular separation is too large, BHs are unable to catch up with their phase delay within the given timeframe.\\

\subsubsection{Black holes from a cosmological sequence of mergers}
\label{satgen}

It is possible to reduce further the parameter set by employing a statistical sampling approach for BH masses, based on a cosmological model. This leaves a single parameter $\Delta \varphi_\mathrm{i}$ assumed to be uniform, allowing the derivation of a probability of capture. The selection of BH masses hinges on the assumption that they originate from the nuclei of two galaxies that underwent major mergers, ultimately forming our target galaxy for study. We also assume that within these nuclei, the BHs were accreting matter from the galaxy, which makes it possible to correlate their mass with that of their host galaxy.\\

\begin{figure}
    \centering
    \includegraphics[width=1.\columnwidth]{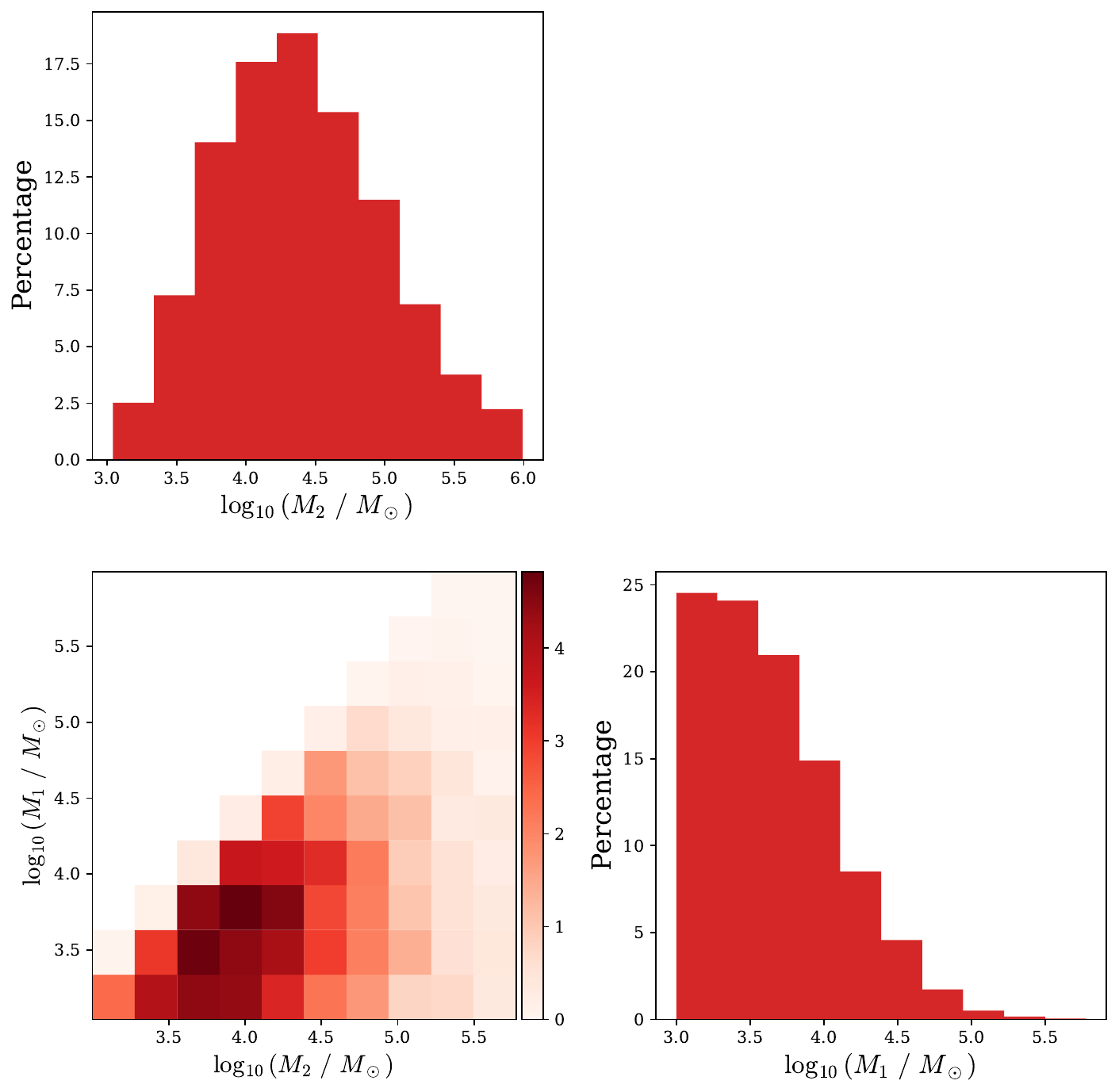}
    \caption{Monte Carlo sampling of BH mass using SatGen.}
    \label{fig-sampleSatgen}
\end{figure}

First, we used dark matter merger trees obtained within the Semi-Analytic Satellite Generator (SatGen) introduced by \citet{satgen}. These merger trees are generated using an algorithm \citep[][]{Parkinson2008} based on the extended Press-Schechter formalism \citep[][]{Lacey93}. We generated $4\,000$ such merger trees for the target halo of mass $M=10^{11} M_\odot$ at z=0, and assess the associated stellar masses of the two progenitors according to the stellar-to-halo mass relation of \citet{Behroozi2013}. Given the 0.16 dex scatter of the relation, we drew 10 realizations of the stellar mass per halo mass, following a lognormal distribution. In the context of accreting BH, mass accumulation is tied to the stellar mass. We assumed that the BH masses follow the $M_{BH}-M_\star$ relation of \citet{Greene2020}\footnote{Low-mass galaxies pose challenges as they lack $\sigma_\star$ measurements and have diminished bulge fractions \citep{MacArthur2003}. Consequently, \citet{Greene2020} revisited the $M_\mathrm{BH}$ -- $M_\star$ correlation, focusing on dynamical studies involving BHs with $M_\mathrm{BH} < 10^6 M_\odot$.}: 
\begin{equation}
    \centering
    \log (M_\mathrm{BH}/\mathrm{M_\odot}) = \alpha + \beta \log (M_\star / M_0) \pm \epsilon
\end{equation}
with $M_0=3\times 10^{10}M_\odot$, $\alpha = 7.43$, $\beta = 1.61$ and $\epsilon = 0.81$ the intrinsic scatter. We drew a series of 10 BH masses per stellar mass assuming again a lognormal distribution given the scatter of the relation. Following a selection process that involves filtering major mergers, requiring progenitors' halo mass ratio to be above $1/3$, we end up with $4\,367$ pairs of BH masses ($M_1, M_2$). A visualization of this distribution is plotted in Fig. \ref{fig-sampleSatgen} and is also reported as contour lines in Fig. \ref{fig-3D}. We then convolved the merger probability (Figure \ref{fig-3D}) with the mass distribution derived from the semi-analytic model (Figure \ref{fig-sampleSatgen}). For each pair of masses ($M_1$, $M_2$) generated by the monte carlo algorithm, we calculated the capture proportion using our results from Section \ref{KaurCase} as

\begin{equation}
    \centering
    p_i(M_1,M_2) = \frac{n(\hbox{bound}\ |\ M_1,M_2)}{n(\hbox{tot}\ |\ M_1,M_2)}
\end{equation}

\noindent with $p_i(M_1, M_2)$ the capture proportion for a given pair of masses, $n(\text{bound} | M_1, M_2)$ the number of captures for the pair of masses ($M_1, M_2$) and $n(\text{tot} | M_1, M_2)$ the total number of simulations conducted for this pair of masses. We finally estimated the total probability:
\begin{equation}
    \centering
    p_{\rm tot} = \frac{1}{N}\sum_{i=1}^N p_i
\end{equation}
where the sum relates to the cases generated by SatGen, $N$ being the number of BH pairs generated. We obtain: $p_{\rm tot} = 13.2\%$. This result, which is not negligible, carries significant implications for the assembly of BH masses, especially if these captures can be sustained over time through dissipative forces. Further investigations into the effectiveness of these forces, particularly within stripped nuclei or globular clusters, are warranted.

\section{Conclusions}
\label{discussion}

\subsection{Summary}

As a result of prolonged dynamical friction timescales and core stalling in cored dark matter haloes, a significant population of off-center intermediate-mass BHs is expected in dwarf galaxies. In this article, we investigate the possibility for such BHs to merge outside the galactic center. Dwarf galaxies being the most common type of galaxy in the Universe, off-center mergers could have a major impact on the assembly of BH mass and gravitational wave emission. We focus on the first step of an off-center merger of IMBHs: Jacobi capture. In order to isolate the capture step, we perform idealized simulations of two BHs in coplanar circular orbits in a smooth background gravitational potential. Given the cored background potential and the fact that the BHs may not be far from the center, the Lagrange points and the zone of influence differ significantly from the Keplerian case (cf. Section \ref{sec-RH}). We find that the captures are highly sensitive to initial conditions, suggesting chaotic dynamics (cf. Section \ref{sec-chaos} and Fig. \ref{fig-chaos}).\\

First in Section \ref{generalCase}, we aim to analyze which configurations, in terms of BH masses and kinematics, favor Jacobi captures. This work allows to pinpoint regions in phase space where Jacobi captures are most likely to occur, paving the way for a comprehensive investigation of the subsequent merger stages through $N$-body simulations. We find that:\\
$\bullet$ In $7.1\%$ of our cases of study (see Table~\ref{tab-5para}), Jacobi captures occur. These captures are categorized into two distinct types: those formed at the beginning of the simulations, referred to as initial captures, and those emerging during the integration at later times, which we call dynamical captures. However, due to the ergodic nature of this study, any stretch of 14 Gyr should be equivalent to another and a time shift in the simulation range would interchange initial and dynamical captures.\\
$\bullet$ Initial captures represent $66.3\%$. They are favored at high radii due to the predominance of the potential component in our capture criterion over energy (Eq. \ref{eq-bindingEnergy}). Conversely, at low radii, the decrease in the circular velocity shear, attributed to the core within the density profile, contributes to their occurrence (see Fig. \ref{fig-iniCapture}).\\
$\bullet$ Dynamical captures make up the remaining $33.7\%$ of the captures. They are favored when the mass of the outer BH exceeds that of the inner one (see Fig. \ref{fig-dynCaptureM1M2}), which is of particular interest within the context of core stalling. We find capture bands in $\Delta r_{\mathrm{i}}$, i.e., a range of radial separation values leading to capture. The capture band moves to higher $\Delta r_{\mathrm{i}}$ as the binary mass increases. To the right of the band ($\Delta r_{\mathrm{i}}$ larger), the BHs are too far away from each other to produce a capture. To the left of the band  ($\Delta r_{\mathrm{i}}$ smaller), dynamical capture is infeasible due to the impossibility of phase catch-up within a Hubble time.\\

Then, in Section \ref{CaptureProb} we reduce the number of free parameters in order to estimate a probability of capture. To reach this goal, we made assumptions about the origin of the BHs and the nature of their shift from the center. \\
$\bullet$ Our first assumption is that BH would be placed on their orbit by core stalling, which fixes their initial radius. We find that in $13.9\%$ of our cases of study (see Table~\ref{tab-3para}), Jacobi captures occur. The increase in the capture percentage compared to the previous scenario suggests that core stalling positions BHs at radii that favor capture.\\
$\bullet$ Given the mass range under study, core stalling positions the BHs within a regime characterized by close radial separations. In this context, the vast majority of captures ($95.4\%$) occur initially, if the BHs are sufficiently close. This condition necessitates $M_1$ and $M_2$ to be close, corresponding to a small radial separation, and a low initial phase shift $\Delta \varphi_\mathrm{i}$ (see Fig. \ref{fig-3D}).\\
$\bullet$ Our second assumption is that these BHs originate from the central regions of two galaxies that underwent a major merger, ultimately forming our target galaxy for study. This enables to recover a statistical distribution of BH masses, employing a Monte Carlo sampling algorithm to construct merger trees and using scaling relations between stellar and BH masses (see Fig. \ref{fig-sampleSatgen}). By doing this, we obtain a probability of capture of $13.2\%$.

\subsection{Caveats}

The primary goal of our work is to gain a better qualitative understanding of Jacobi capture on a case-by-case basis. On the way, we had to make two particular assumptions that will need to be alleviated in the future, as they may significantly impact our picture for BH growth: we restricted the problem to specific initial orbits, and we assumed a static galactic potential.

The first restrictive assumption on BH orbits concerns the geometry of the problem. Indeed, we considered only coplanar BHs, while in all generality their orbits could be randomly oriented. It may seem that this assumption artificially boosts our probability of Jacobi capture, because coplanar orbits imply that the BHs are, on average, closer both spatially, and in velocity. However, in the scenario where off-center BHs originate from accretion events, it is expected that such occurrences demonstrate spatial correlation. Consequently, the BH orbital planes have correlated orientations, in a way that favors coplanar orbits. The second assumption on the BH orbits is that they are initially circular. But this is actually quite realistic, because dynamical friction tends to circularize a wandering BH's orbit, all the more that it has been orbiting for a long time. Finally, we assumed all orbits to revolve in the same sense, which effectively boosts the probability of capture by a factor 2. Similarly, this factor could be mitigated by the correlated origin of accreted BHs.

The assumption of a static background galaxy may also be a concern. Indeed, in some of our cases of study, we expect the reflex response from the galaxy to be significant. This is particularly the case for a stalling BH, where the very definition of stalling implies that the galaxy is responding. This motion of the background may induce intricate modifications of the dynamics of individual setups, even at the level of a single BH orbiting a galaxy. The consequences could be even stronger in the case of two BHs. Then, dynamical friction and buoyancy exerted on each individual BH could couple, in a way that would even modify the stalling radii of each BH, or suppress the stalling process altogether. A more detailed treatment of such cases with live $N$-body runs is required to assess the influence of the galactic response on Jacobi capture.

In Section \ref{satgen} we have selected the BH masses from a semi-analytical model, it is important to note that it is contingent upon the assumption that they are remnants of central BHs from past major mergers. The application of a correlation between stellar mass and BH mass hinges on the assumption of BH accretion of galactic matter. For scenarios involving wandering BHs situated outside the galactic center, refraining from accretion, reference to Fig. \ref{fig-3D} is required on a case-by-case basis for each BH mass pair.

\subsection{Implications for black holes}

In this study we investigate if and under which circumstances two BHs can become bound while outside the center of a galaxy. We do find that such captures do happen at substantial fractions in our simplified simulations. In recent years, IMBHs were successfully detected outside the centers of galaxies \citep{Reines2020}, as well as single relic nuclear clusters at kpcs from the center of the galaxy have been confirmed \citep{Seth2014, Ahn2018}.  Simulations and model predictions show that these relics of the accretion event should be extremely common \citep{Tremmel2018, Voggel2019} and mergers have been shown to be the reason for the M-$\sigma$ relation \citep{Volonteri2009}.

If the hidden off-center population of BHs merge repeatedly before reaching the center they would increase their mass with smaller merging events and be more massive once they reach the center of the main galaxy. The existence of off-center BH mergers also implies that three massive BHs simultaneously descending toward the galaxy's center becomes more rare. This would cause an unstable three-body system ensues, which can lead to the ejection of one BH. However, if two of the three BHs merge during their inspiral path, only two BHs reach the center, enabling a more stable situation and thus a more likely subsequent merger that yields a more massive BH than in the previous scenario. The history of BH mergers being altered by the possibility of off-center mergers, this should change the expected gravitational wave signals.

As this type of mergers can have major implications on the mass growth of BHs but has not been explored much in the literature it is crucial to improve upon our idealized simulations. The next step in the investigation of the off-center merger phenomen could be to determine whether the inclusion of stellar populations associated with BHs, such as stripped galactic nuclei or stars within globular clusters changes the Jacobi captures. Such an inclusion is necessary to be more realistic, as typically BHs are surrounded by nuclear star clusters \citep{Seth2008}. In particular it will be important to see whether dissipative forces such as dynamical friction between the stellar populations can help sustain these captures.


%

\begin{acknowledgements}

    We thank Rapha\"el Errani, Florent Renaud, Paolo Bianchini, Katarina Kraljic, Jorge Pe\~narrubia, Anna Lisa Varri, Marta Volonteri, Benoit Famaey, Gary Mamon, Karamveer Kaur and Julien Montillaud for the helpful discussion and the anonymous referee for the constructive suggestions.

    The authors would like to acknowledge the High Performance Computing Center of the University of Strasbourg for supporting this work by providing scientific support and access to computing resources. Part of the computing resources were funded by the Equipex Equip@Meso project (Programme Investissements d'Avenir) and the CPER Alsacalcul/Big Data.
    
    This research made use of the following software: \texttt{Agama} \citep[][]{Vasiliev2018}, \texttt{Jupyter} \citep{jupyter}, \texttt{Matplotlib} \citep{Matplotlib}, \texttt{Numpy} \citep{Numpy}, \texttt{Python} \citep[][]{python}, \texttt{SatGen} \cite[][]{satgen} and \texttt{Scipy} \citep[][]{scipy}.
    Simulations in this paper made use of the REBOUND N-body code \citep{rebound}. The simulations were integrated using IAS15, a 15th order Gauss-Radau integrator \citep{reboundias15}.
  
\end{acknowledgements}

%
%

\bibliographystyle{aa} 
\bibliography{biblio.bib} 

\begin{appendix} 

\section{Binary eccentricity during Jacobi captures} \label{sec-ecc}

\begin{figure}[h!]
    \centering
    \includegraphics[width=1.\columnwidth]{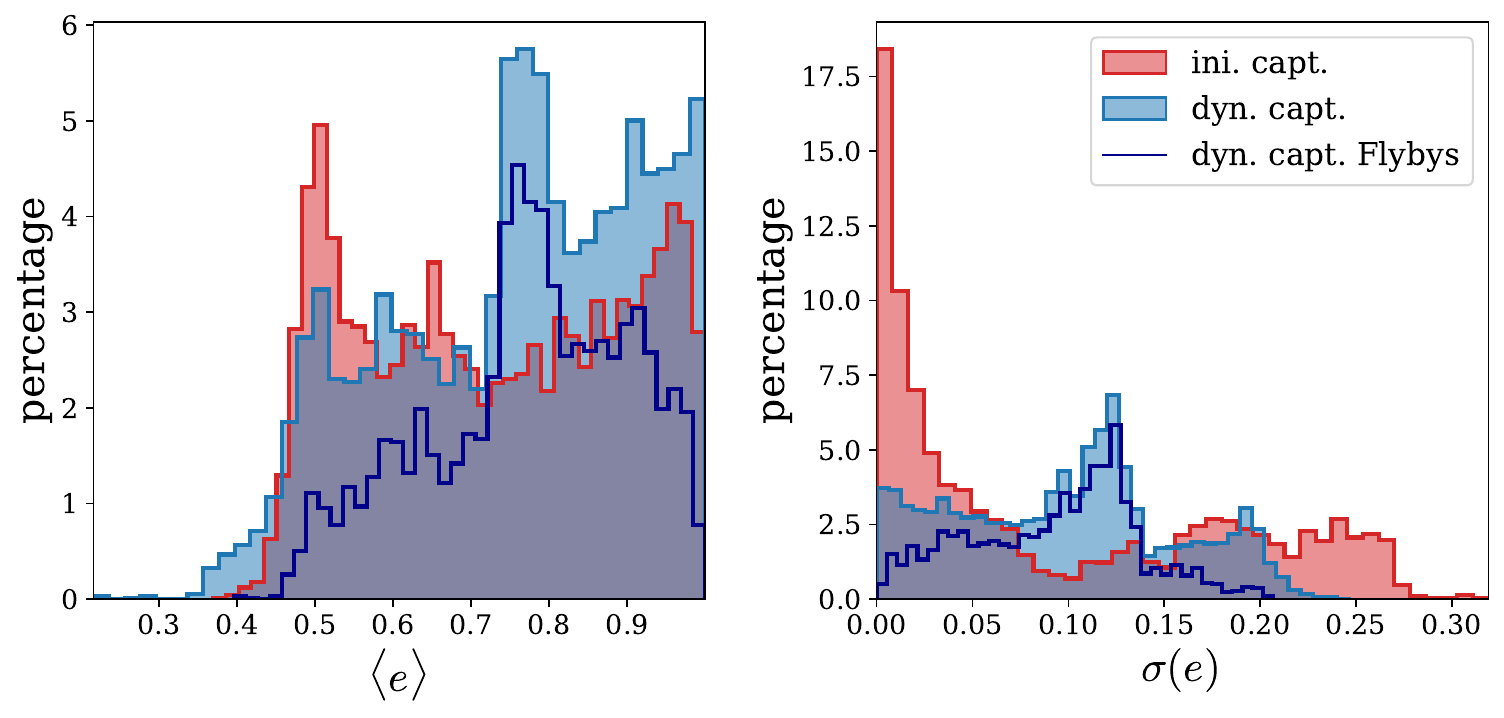}
    \caption{Eccentricity during Jacobi captures. Left panel: histogram in percent of the mean eccentricity of the binary during captures (initial captures in red and dynamical captures in blue). The dark blue line represents only dynamical captures with a single close encounter (i.e., flybys). Jacobi captures produce rather eccentric binaries. A significant proportion are in the very high eccentricity range ($e \gtrsim 0.9$), for which Post-Newtonian corrections can be significant at the pericenter. Right panel: percentage histogram of standard deviation of eccentricity during captures. These variations may permit a very high eccentricity during the capture, despite the lower mean eccentricity.}
    \label{fig-ecc}
\end{figure}

The histogram in the left panel of Fig. \ref{fig-ecc} displays the mean eccentricity of the binary during captures. Jacobi captures tend to result in rather eccentric binaries, a factor that can aid in stabilization through energy dissipation caused by dissipative forces. Post-Newtonian corrections can particularly influence highly eccentric orbits, especially at the pericenter \citep{Samsing2018}. Notably, our criterion ensures that the mean eccentricity during captures remains below 1. On the right panel of Fig. \ref{fig-ecc}, the standard deviation of eccentricity during captures is depicted. These variations may permit a very high eccentricity during the capture, despite the lower mean eccentricity.

\section{Alternative capture criterion} 
\label{sec-crit}

In this section, we revisit some of the findings presented in the article using an alternative capture criterion based solely on the radius of influence (i.e., only on gravitational potential). Here, we define BHs as captured when their separation is less than the Hill radius of the binary. This approach allows us to emphasize the role of the kinetic component of the criterion utilized throughout the paper (Eq. \ref{eq-bindingEnergy}).\\

When the separation between the BHs is smaller than the binary Hill radius, the potential generated by the BHs competes with that of the galaxy, resulting in a significant perturbation of the BH trajectories. Therefore, it is natural to consider that this criterion alone may suffice to define a capture. We examine here the differences that this approach entails.\\

Applying this new criterion to the general five-parameter case, we find captures in $16.4\%$ of the simulated instances. This is more than double the $7.1\%$ obtained with the energy criterion. There are $5.4\%$ initial captures (compared to $4.7\%$ with the energy criterion) and $11.0\%$ dynamical captures (compared to $2.4\%$). We observe a significant increase in the number of dynamical captures.\\

\begin{figure}[h!]
    \centering
    \includegraphics[width=1.\columnwidth]{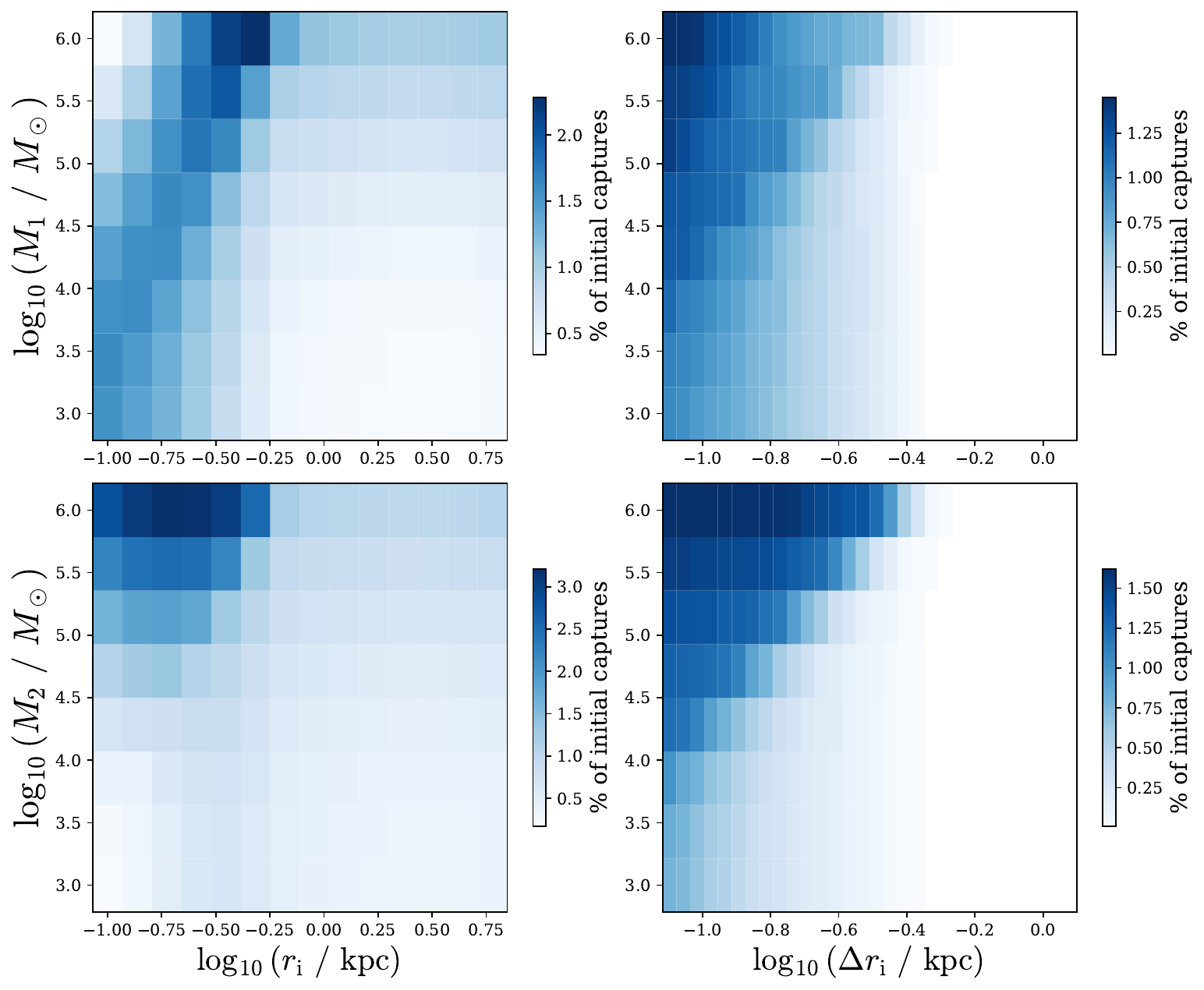}
    \caption{Percentage of initial captures as a function of different parameter pairs, marginalising each time over the three others. Top-left panel: as a function of the mass of the inner BH ($M_1$) and its initial radius ($r_\mathrm{i}$); bottom-left panel: as a function of the mass of the outer BH ($M_2$) and $r_\mathrm{i}$; top-right panel: as a function of $M_1$ and the initial radial separation between the BHs ($\Delta r_\mathrm{i}$); bottom-right panel: as a function of $M_2$ and $\Delta r_\mathrm{i}$. This figure is analogous to Fig. \ref{fig-iniCapture} with a different capture criterion. BHs are considered bound if their relative distance falls below the binary's Hill radius.}
    \label{fig-IniOnlyRH}
\end{figure}

Figure \ref{fig-IniOnlyRH} serves as the counterpart to Fig. \ref{fig-iniCapture}, depicting the percentage of initial captures across various parameter pairs, marginalized over the remaining three. Unlike the latter, this figure does not consider the relative velocity between BHs. In the top-left panel, the percentage of captures with respect to $r_\mathrm{i}$ closely mirrors the Hill radius curve depicted in Figure \ref{fig-RH}, albeit with a leftward shift as the BH mass decreases.\\

The white area in the top-left corner of Fig. \ref{fig-IniOnlyRH} appears dark on the figure \ref{fig-iniCapture} due to cases where the BHs initially maintain a relative distance slightly larger than the Hill radius, yet exhibit very low relative velocities. They are energetically close and plunge into the Hill sphere after a few time steps. Under this criterion, they are classified as dynamical captures, while their low relative velocity categorizes them as initial captures according to the energy criterion.

\begin{figure}[h!]
    \centering
    \includegraphics[width=1.\columnwidth]{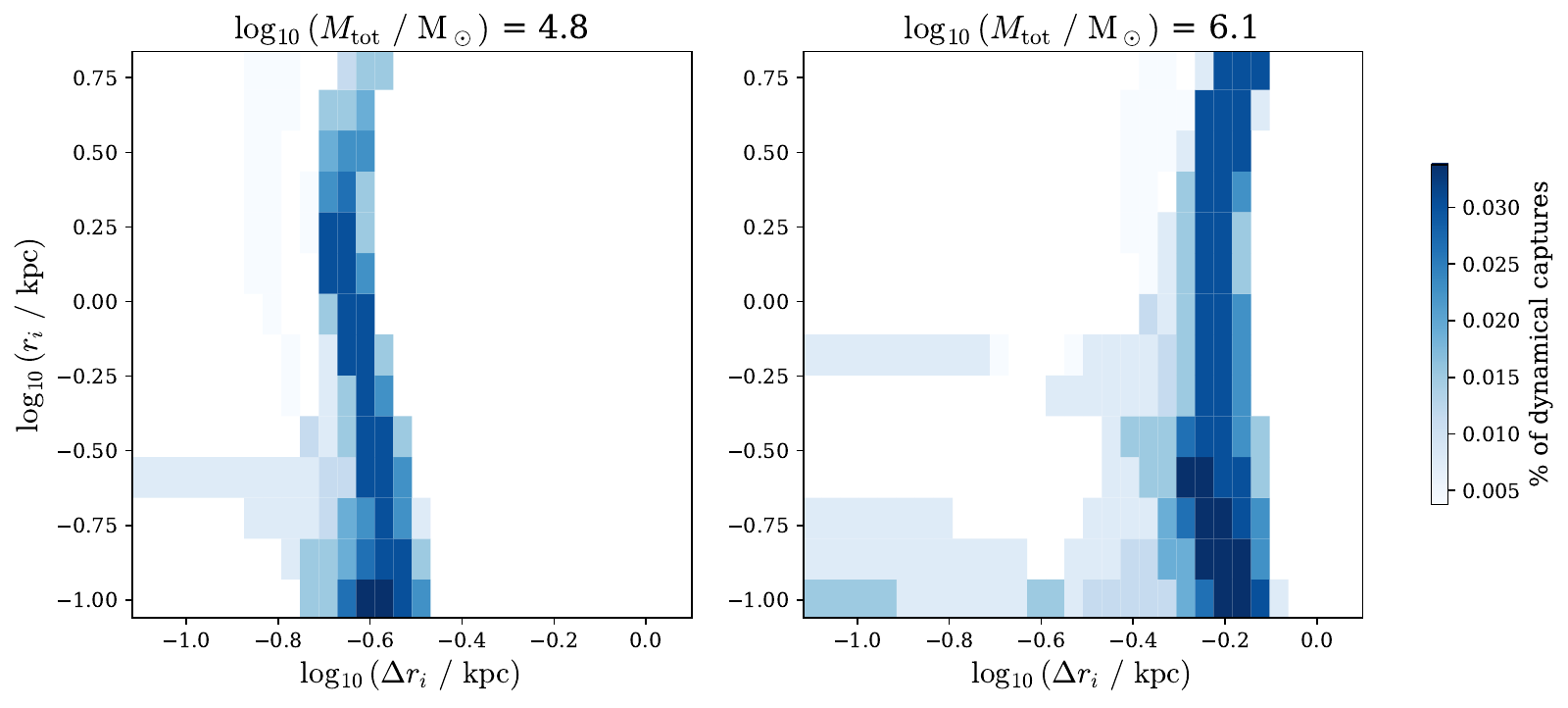}
    \caption{Percentage of dynamical captures as a function of initial inner BH radius ($r_{\mathrm{i}}$) and initial radial separation ($\Delta r_{\mathrm{i}}$) for two mass pairs, On the left panel, $\log (M_1/M_\odot) = 3.9$ and $\log (M_2/M_\odot) = 4.7$, on the right panel, $\log (M_1/M_\odot) = 5.6$ and $\log (M_2/M_\odot) = 6$. This figure is analogous to Fig. \ref{fig-frangesColor} with a different capture criterion. BHs are considered bound if their relative distance falls below the binary's Hill radius. We observe that the binding bands of dynamical captures extend closer to the galactic center. The majority of these captures involve hyperbolic flybys, indicating that the BHs involved have high relative velocities.}
    \label{fig-DynOnlyRH}
\end{figure}

\begin{figure}[h!]
    \centering
    \includegraphics[width=1.\columnwidth]{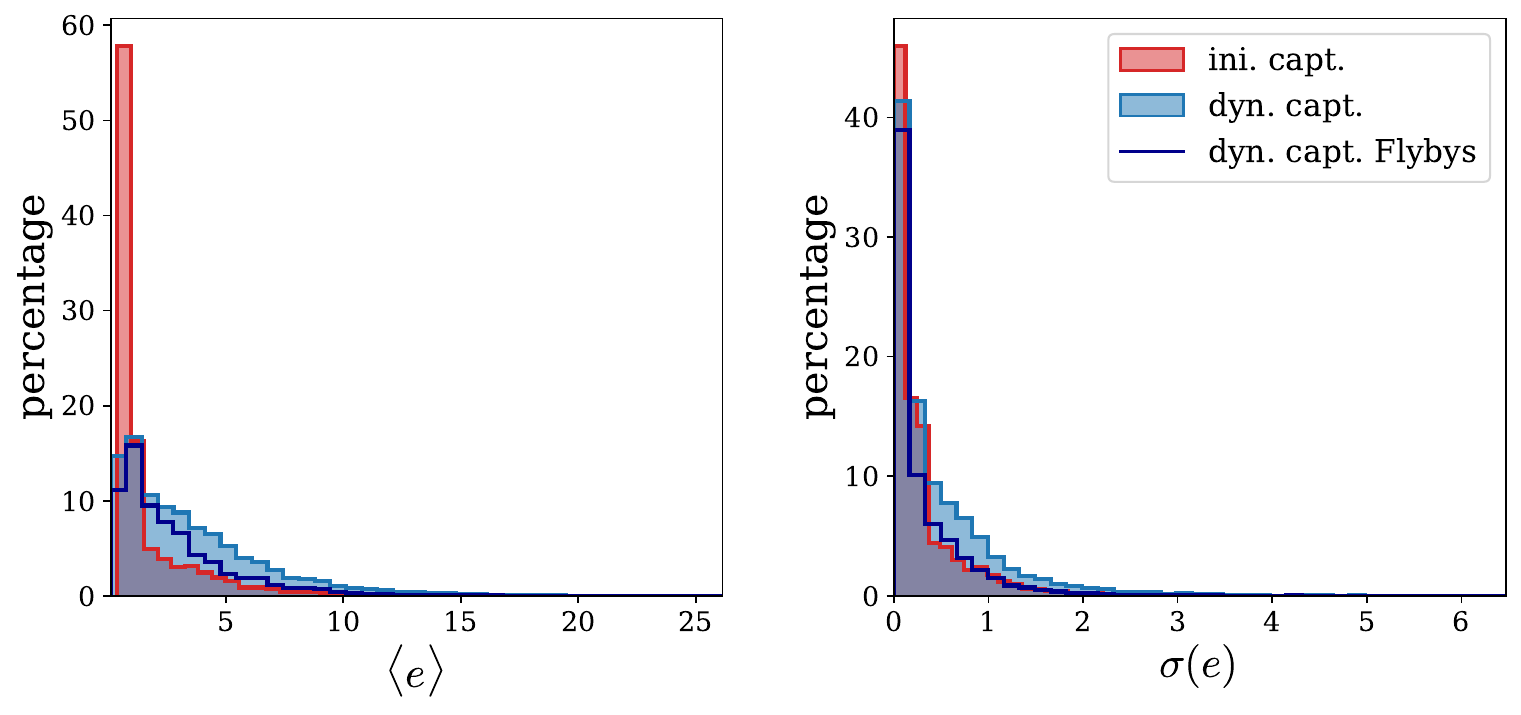}
    \caption{Eccentricity during Jacobi captures. Left panel: histogram in percent of the mean eccentricity of the binary during captures (initial captures in red and dynamical captures in blue). The dark blue line represents only dynamical captures with a single close encounter (i.e., flybys). Right panel: percentage histogram of standard deviation of eccentricity during captures. This figure is analogous to Fig. \ref{fig-ecc} with a different capture criterion. BHs are considered bound if their relative distance falls below the binary's Hill radius. Note that the alternative criterion adds captures for which the mean eccentricity is greater than 1.}
    \label{fig-eccOnlyRH}
\end{figure}

Figure \ref{fig-DynOnlyRH} serves as the counterpart to Fig. \ref{fig-frangesColor}, depicting the percentage of dynamical captures relative to the initial radius of the inner BH ($r_\mathrm{i}$) and the initial radial separation between the BHs ($\Delta r_\mathrm{i}$). Comparing it with Figure \ref{fig-frangesColor}, we observe that the dynamical binding bands extend further down in $r_\mathrm{i}$. To understand the nature of these dynamical captures introduced by the alternative criterion, we can turn to Fig. \ref{fig-eccOnlyRH}, mirroring Figure \ref{fig-ecc}. The histogram of mean eccentricities during captures on the left-hand panel extends beyond 1. Consequently, these newly added dynamical captures are mostly hyperbolic flybys close to the galactic center and for which BHs have high relative velocities.

\section{Pacman orbit and number of close encounters} 
\label{sec-pacman}

\begin{figure}[h!]
    \centering
    \includegraphics[width=.8\columnwidth]{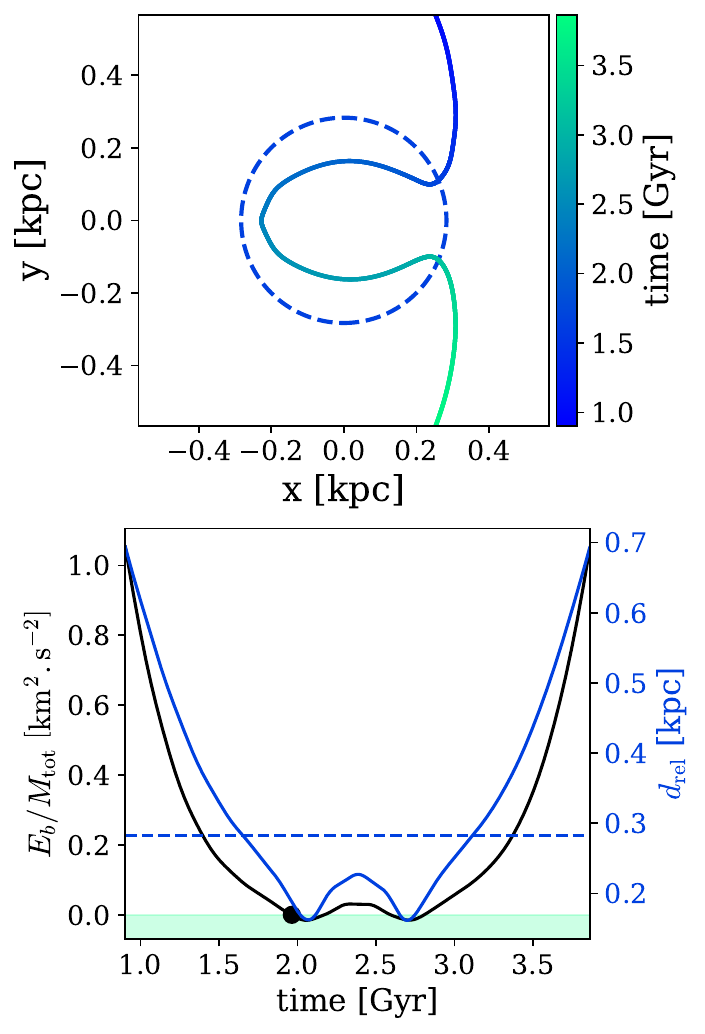}
    \caption{Example of Pacman orbit. Top row: relative coordinates between BHs during the Jacobi capture. The reference frame is centered on the most massive BH and rotates with it. The binary Hill radius is marked by the dotted circle. Bottom row: the black line represents the binding energy ($E_\mathrm{b}$) per unit mass of the binary ($M_\mathrm{tot}=M_1+M_2$), with negative values shown in the colored area. The black marker indicates the point when the energy becomes negative. The blue curve represents the relative distance ($d_\mathrm{rel}$) between the BHs. The binary Hill radius is indicated by a dashed blue line.}
    \label{fig-pacman}
\end{figure}

The existence of a central core within the density profile facilitates the emergence of a novel orbital type known as Pacman orbits \citep{Banik2022}. These orbits play a key role in the phenomenon of dynamical buoyancy within cores, which drives perturbers outward to an equilibrium radius with the dynamical friction, ultimately resulting in core stalling. During capture events, BHs experience these unique dynamics. Figure \ref{fig-pacman} illustrates an example of a Pacman capture.\\

This example serves as a perfect illustration of how the labeling of the number of close encounters between BHs depends, in some borderline cases like this one, on the criterion used to label a capture. In the bottom panel of Fig. \ref{fig-pacman}, the binding energy and relative distance between the BHs during the capture are depicted. With the energy-based capture criterion (Eq. \ref{eq-bindingEnergy}), we would label two captures with a single close encounter, as the energy returns to positive values between the encounters. However, if we utilize a criterion solely based on distance, as outlined in Section \ref{sec-crit} (where BHs are labeled as captured when the relative distance is less than the binary Hill radius), we would label only a single capture with two close encounters. This case underscores the complexity of selecting an appropriate capture criterion, as finding a perfect one remains a challenging task.

\end{appendix}
\end{document}